\documentclass[twocolumn]{aastex63}

\usepackage{natbib}
\usepackage[]{threeparttable}
\usepackage[caption = false]{subfig}
\usepackage{graphicx}

\usepackage{rotating}      
\usepackage{mathrsfs}
\usepackage{array}

\usepackage{color}
\usepackage{amsmath}

\usepackage{multirow,tabularx}
\usepackage{calc}

\newsavebox{\twofigures}

\usepackage{silence}
\graphicspath{{./}{figures/}}

\shorttitle{New NGC~5128 globular clusters at large radii}
\shortauthors{Hughes et al.}

\begin{document}

\title{ New Velocity Measurements of NGC~5128 Globular Clusters out to 130 kpc: Outer Halo Kinematics, Substructure and Dynamics \footnote{This paper includes data gathered with the 6.5 m Magellan Telescope at Las Campanas Observatory, Chile.}}

\correspondingauthor{Allison K. Hughes}
\email{akhughes@email.arizona.edu}

\author[0000-0002-1718-0402]{Allison K. Hughes}
\affil{Steward Observatory, University of Arizona, 933 North Cherry Avenue, Tucson, AZ 85721, USA}

\author[0000-0003-4102-380X]{David J. Sand}
\affil{Steward Observatory, University of Arizona, 933 North Cherry Avenue, Tucson, AZ 85721, USA}

\author[0000-0003-0248-5470]{Anil Seth}
\affiliation{Department of Physics \& Astronomy, University of Utah, Salt Lake City, UT, 84112, USA}

\author[0000-0002-1468-9668]{Jay Strader}
\affiliation{Center for Data Intensive and Time Domain Astronomy, Department of Physics and Astronomy, Michigan State University, East Lansing, MI 48824, USA}

\author[0000-0003-1731-0497]{Chris Lidman}
\affiliation{Centre for Gravitational Astrophysics, College of Science, The Australian National University, ACT 2601, Australia}
\affiliation{The Research School of Astronomy and Astrophysics, Australian National University, ACT 2601, Australia}

\author[0000-0001-6215-0950]{Karina Voggel}
\affiliation{Universite de Strasbourg, CNRS, Observatoire astronomique de Strasbourg, UMR 7550, F-67000 Strasbourg, France}

\author[0000-0003-0234-3376]{Antoine Dumont}
\affiliation{Department of Physics \& Astronomy, University of Utah, Salt Lake City, UT, 84112, USA}

\author[0000-0002-1763-4128]{Denija Crnojevi\'{c}}
\affiliation{University of Tampa, 401 West Kennedy Boulevard, Tampa, FL 33606, USA}

\author{Mario Mateo}
\affiliation{Department of Astronomy, University of Michigan, Ann Arbor, MI 48109, USA}

\author[0000-0003-2352-3202]{Nelson Caldwell}
\affiliation{Center for Astrophysics, Harvard \& Smithsonian, 60 Garden Street, Cambridge, MA 02138, USA}

\author{Duncan A. Forbes}
\affiliation{Centre for Astrophysics and Supercomputing, Swinburne University of Technology, Hawthorn VIC 3122, Australia}

 \author[0000-0003-0256-5446]{Sarah Pearson}\thanks{Hubble Fellow}
\affiliation{Center for Cosmology and Particle Physics, Department of Physics, New York University, 726 Broadway, New York, NY 10003, USA}

\author[0000-0001-8867-4234]{Puragra Guhathakurta}
\affiliation{UCO/Lick Observatory, University of California Santa Cruz, 1156 High Street, Santa Cruz, CA 95064, USA}

\author[0000-0001-6443-5570]{Elisa Toloba}
\affiliation{Department of Physics, University of the Pacific, 3601 Pacific Avenue, Stockton, CA 95211, USA}

%%%%%%%%%%%%%%%%%%%%%%%%%%%%%%%%%%%%%%%%%%%%%%%%%%%%%%%%%%%%%%%%%%%%%%%%%%%%%%%%
\begin{abstract}

We present new radial velocity measurements from the Magellan and the Anglo-Australian Telescopes for 174 previously known and 122 newly confirmed globular clusters (GCs) around NGC 5128, the nearest accessible massive early-type galaxy at D=3.8 Mpc. Remarkably, 28 of these newly confirmed GCs are at projected radii $> 50\arcmin$\ ($\gtrsim 54$ kpc), extending to $\sim 130$ kpc, in the outer halo where few GCs had been confirmed in previous work. We identify several subsets of GCs that spatially trace halo substructures that are visible in red giant branch star maps of the galaxy. In some cases, these subsets of GCs are kinematically cold, and may be directly associated with and originate from these specific stellar substructures. From a combined kinematic sample of 645 GCs, we see evidence for coherent rotation at all radii, with a higher rotation amplitude for the metal-rich GC subpopulation. Using the tracer mass estimator, we measure a total enclosed mass of $2.5\pm0.3 \times 10^{12} M_{\odot}$ within $\sim 120$ kpc, an estimate that will be sharpened with forthcoming dynamical modeling. The combined power of stellar mapping and GC kinematics makes NGC~5128 an ongoing keystone for understanding galaxy assembly at mass scales inaccessible in the Local Group.

\end{abstract}
 
%Optical identification (1167), Galaxy evolution (594)}
% see here: https://journals.aas.org/keywords-2013/
%%%%%%%%%%%%%%%%%%%%%%%%%%%%%%%%%%%%%%%%%%%%%%%%%%%%%%%%%%%%%%%%%%%%%%%%%%%%%%%%


\defcitealias{Hughes2021}{H21}

\defcitealias{Woodley2010a}{W10}

\section{Introduction}

Globular clusters (GCs), compact massive star clusters found in substantial numbers in all massive galaxies, provide a window into the various epochs of star formation that mark a galaxy's evolutionary history. In nearby galaxies ($D \lesssim 20$ Mpc), where detailed investigations are possible, GC systems can provide important insights into the chemical and dynamical history of their hosts. 
Due to their high luminosities and compact sizes, GCs are observed much more easily than the underlying stellar field components in the remote parts of galaxies, allowing efficient photometry and spectroscopy. Various studies have found correlations between GC systems and their host galaxy properties that can shed light on galaxy formation mechanisms \citep{Brodie06}. GC kinematics provide information about the assembly history of the host galaxy, its total mass, and its dark matter distribution (e.g., \citealt{Schuberth2010, Schuberth2012, Strader2011,Alabi17}). GCs also provide an alternative way to look for and study past accretion events, by searching for spatially and dynamically linked GC groups that can serve as tracer populations for their (now disrupted) parent systems (e.g., \citealt{Mackey10,Veljanoski2014,Kirihara2017,Mackey2019}). 

The most numerous populations of GCs are found in luminous elliptical galaxies, especially those at the center of massive groups or clusters \citep{Harris1979,Brodie06,Richtler2006}, which can have thousands of GCs. However, there are still only a handful of galaxies that have a large population ($\gtrsim 500$) of GCs with measured radial velocities, including the cluster-central galaxies M87 \citep{Cote2001, Strader2011,2015ApJ...802...30Z,Forbes17} and NGC~1399 \citep{Richtler2004,2010A&A...513A..52S,2018MNRAS.481.1744P}. These large samples of velocities are necessary to move beyond rough halo mass estimates to measurements of the halo concentration, searches for substructure, and constraints on the ongoing assembly of the dynamically young outer regions of the halo (e.g., \citealt{2012ApJ...748...29R,2022A&A...657A..93C}).

Historically, NGC~5128 (Centaurus A) has been a leading target for extragalactic GC studies, due to its proximity and the richness of its GC system.  NGC~5128 is the central elliptical galaxy in a group of galaxies at a distance of $3.8 \pm 0.1$ Mpc \citep{Harris2010}. Since the first discovery of GCs in NGC~5128 in the 1980s, many photometric and spectroscopic surveys have been conducted, leading to the identification of $\sim$600 confirmed GCs and thousands of GC candidates \citep{Graham1980, VanDenBergh1981, Hesser1986, Harris1992,  Harris2004,  Harris2012, Holland1999, Harris2002, Harris2006, Peng2004gc1, Martini2004,   Woodley2005, Woodley2007, Woodley2010a, Woodley2010b, Gomez2006,  Rejkuba2007, Beasley2008, Georgiev2009,  Georgiev2010, Taylor2010, Taylor2015, Taylor2017, Mouhcine2010, Sinnott2010,  Voggel2020, Fahrion2020, Muller2020, Dumont2021}. While the bulk of the stellar mass in NGC~5128 is that of an old metal-rich massive elliptical galaxy, the galaxy also shows a peculiar shape and  recent star formation that point to a relatively recent gas-rich merger, as well as numerous halo substructures that point to an ongoing active accretion history (e.g., \citealt{Baade1954, Graham1979, Israel1998, Peng2004gc2, Crnojevic16,Wang2020}). 

The goal of the present study is to improve our understanding of the outer halo of NGC~5128 via a newly enlarged, updated sample of radial velocity-confirmed GCs. We combine these GC velocities with a wide-field resolved red giant branch (RGB) star map of the galaxy out to a projected galactocentric radius of $\sim$ 150 kpc produced by the Panoramic Imaging Survey of Centaurus and Sculptor (PISCeS;  \citealt{Crnojevic16}), allowing matches between outer halo GCs and specific stellar substructures. We show that we can successfully apply this method of tracing individual galactic accretion signatures with GCs to NGC~5128, which had previously only been used on a large scale inside the Local Group (e.g., \citealt{Veljanoski2014}).

This paper is organized as follows.
Section~\ref{sec:obs_reduc} describes our observations and data reduction process for optical spectroscopy from Magellan/M2FS and AAT/AAOmega.
In Section~\ref{sec:radial_velocities} we discuss our methodology for measuring radial velocities and present our sample of radial velocity measurements for previously known and newly confirmed GCs in NGC~5128. 
We analyze the updated GC population in Section~\ref{sec:analysis}. In Section~\ref{sec:GCsub} we identify GCs  belonging to specific substructures within the halo of NGC~5128, and estimate a mass profile for the galaxy in Section~\ref{sec:total_mass}.
 We summarize and conclude in Section~\ref{sec:conclude3}.

As in our previous GC work around NGC~5128 in \cite{Hughes2021}, hereafter \citetalias{Hughes2021},  we adopt a distance modulus for NGC~5128 of
$(m - M)_{0} = 27.91$ mag, corresponding to a distance of $D =3.82$ Mpc \citep{Harris2010}. The physical scale at this distance is 18.5 pc~arcsec$^{-1}$ (1.1 kpc~arcmin$^{-1}$).

%%%%%%%%%%%%%%%%%%%%%%%%%%%%%%%%%%%%%%%%%%%%%%%%%%%%%%%%%%%%%%%%%%%%%%%%%%%%%%%%
% Data Sets
%%%%%%%%%%%%%%%%%%%%%%%%%%%%%%%%%%%%%%%%%%%%%%%%%%%%%%%%%%%%%%%%%%%%%%%%%%%%%%%%
\section{Observations and data reduction}\label{sec:obs_reduc}

In this section, we first summarize our GC selection technique, underpinned by our work presented in \citetalias{Hughes2021}.  Following this, we present our new Magellan/M2FS and AAT/AAOmega optical spectroscopy to identify new radial velocity confirmed GCs in the NGC 5128 system out to large radii.

\subsection{GC candidate selection}

\citetalias{Hughes2021} focused on GC selection out to a projected radius of $\sim 150$ kpc from the galaxy center using the 95 PISCeS fields \citep{Crnojevic16,Crnojevic19}. To summarize the steps of this process: (i) using the newly derived photometry, we first made a selection in magnitude, excluding both the brightest GC candidates ($r\lesssim 18$ mag, affected by saturation) and faintest ones ($r\gtrsim 22$ mag, beyond which ancillary data is mostly missing and contamination is substantial), as well as those GCs affected by crowding near the galaxy's center ($<$10\arcmin\ from NGC~5128); (ii) we did a likelihood-based selection of extended objects likely to be GCs using a two-aperture technique on the PISCeS photometry, measured by a concentration index $C_{3-6}$ (the difference in magnitude between 3 and 6 pixel apertures); (iii) we rejected background galaxies and stellar blends 
using cuts in the effective radius, ellipticity, and large-radius flux distribution; (iv) we added in data from Gaia DR2 to reject foreground stars using measurements of astrometric motion (proper motion and parallax) and proxies for extendedness (astrometric excess noise and $BP$/$RP$ excess), as GCs at the distance of NGC~5128 are slightly resolved; (v) we used multi-band photometry from the NOAO Source Catalog (NSC) to assign a likelihood that each source has colors consistent with known GCs. For each step a quantitative likelihood was assigned, and a final \textit{total\_likelihood} for each candidate was calculated by multiplying each of these likelihoods together (with no penalty when data was missing).  Numbers closer to unity represent a larger chance that the source is a GC associated with NGC~5128. Again, we refer the reader to \citetalias{Hughes2021} for details.

Based on the amount of data available, each GC was also placed into one of four categories: \emph{gold} if the GC had data in PISCeS, Gaia DR2, and NSC; (2) \emph{silver} if the GC had data in PISCeS and one of Gaia DR2 or NSC; (3) \emph{bronze} if the GC had data only in PISCeS and was well-resolved with $C_{3-6} > 2.0$; and (4)  \emph{copper} if the GC had data only in PISCeS and was marginally resolved with $2.0 > C_{3-6} > 1.0$. We identified a total of 40502 GC candidates, the vast majority of which are likely contaminants. Of these, we highlighted the 1931 gold and silver candidates with \textit{total\_likelihood}$>$0.85 as the highest priority for spectroscopic follow-up for confirmation (see Table 3 in \citetalias{Hughes2021} for a breakdown of the number of GC candidates in each data rank).

The GC candidate selection process described above was designed for GCs in the magnitude range 18 $\lesssim$ $r$ $\lesssim$ 22 mag, and does not include either very bright or very faint GCs. For bright GCs, our team used a  parallel effort to identify bright GCs/UCDs ($L_V \geq 2 \times 10^5~L_\odot$, roughly corresponding to $M \geq 3 \times 10^5~M_\odot$) out to similarly large radii ($\sim 150$ kpc) \citep{Voggel2020,Dumont2021}. Also,  beyond the conservative color selection described above,  we did not further prioritize between red and blue GC candidates, and so there should be no preferential selection of metal-rich versus metal-poor GCs. Hence this process should lead to a minimally biased sample of candidate GCs with $r \lesssim 22$ in NGC 5128. No matter the selection method, we also include all known, radial velocity confirmed GCs in our kinematic analyses in later sections of this work.

The optical spectroscopy detailed below was taken over five years, with some of the earliest data obtained before our final rigorous likelihood-based target selection was in place and hence based on preliminary photometry with less ancillary data. We evaluate the in-practice success of our selection methods in Section \ref{sec:roi}.

\begin{figure}[t]
\includegraphics[width=1.0\linewidth]{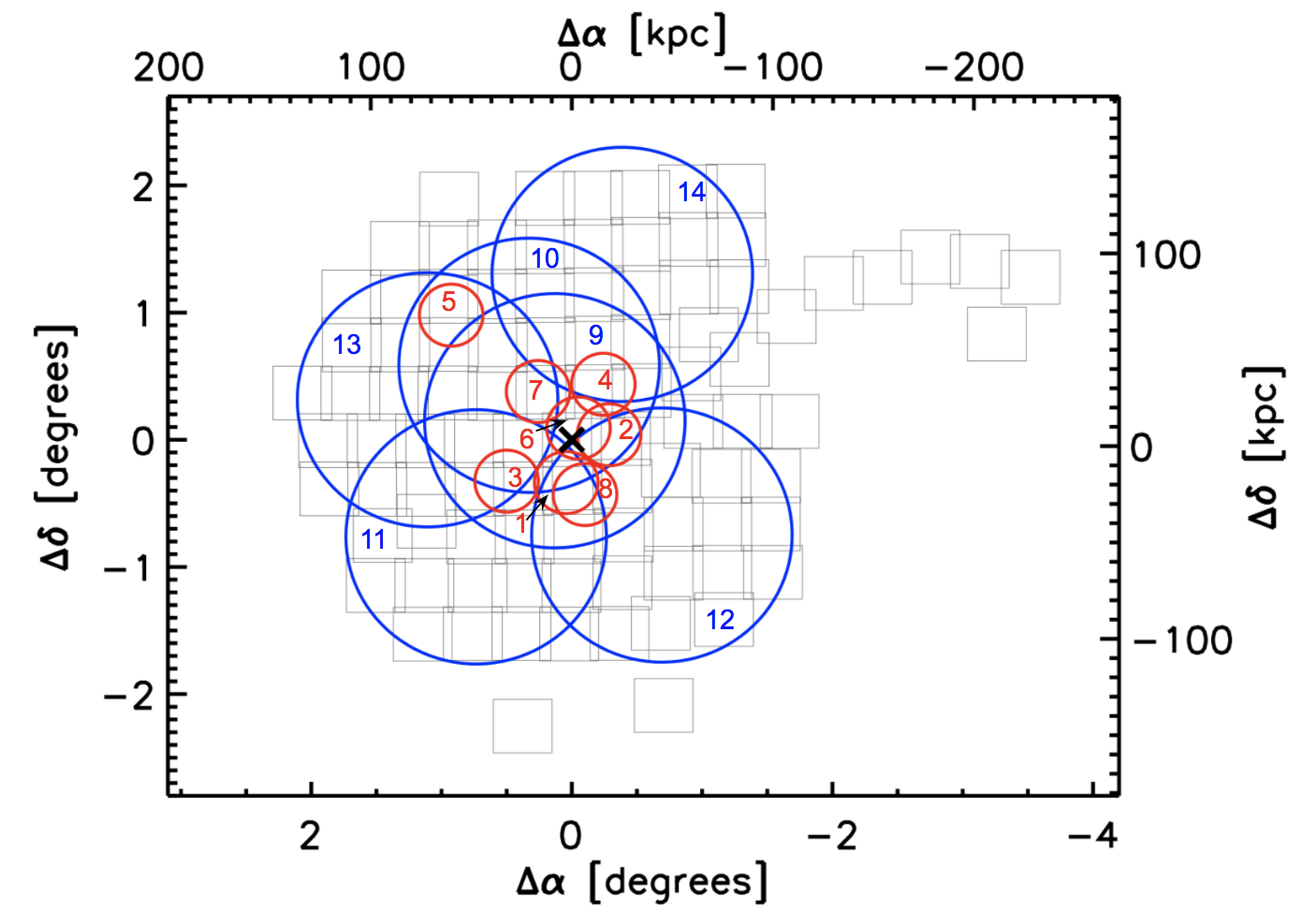}
\caption{Footprint of the PISCeS survey (small pale squares) around NGC~5128, oriented such that north is up and east is left.  The center of NGC~5128, located at $\alpha$=$201.362540^{\circ}$, $\delta$=$-43.033627^{\circ}$, is marked by the central black `x'. Regions observed with Magellan/M2FS are indicated by the red circles and those observed with AAT/AAOmega are indicated by the blue circles, with positions given in Tables \ref{table:m2fs_observing} and \ref{table:2df_observing}.}
\label{fig:map_observed}
\end{figure}

\subsection{Magellan/M2FS}\label{sec:m2fs_data}

We obtained multi-object fiber spectroscopy using the Michigan/Magellan Fiber System (M2FS) instrument on the Magellan/Clay 6.5 m telescope at the Las Campanas Observatory on several runs between 2017 and 2019.
The M2FS Spectrograph consists of twin spectrographs, each of which can be fed by up to 128 1.2\arcsec\ diameter fibers over a field of view 30\arcmin\ in diameter \citep{Mateo2012}.  
The twin spectrographs can be set to different resolution modes, and we used this to accomplish two science goals simultaneously.  
The ``high" resolution echelle mode was used to measure stellar velocity dispersions of bright globular clusters and ultra-compact dwarfs to identify stripped galaxy nuclei that might contain massive central black holes, as discussed in \cite{Dumont2021}.  In the current work, we use the ``low" resolution data obtained from the other spectrograph, with a resolution of $R\sim1300$ and wavelength range 4100--6000 \AA, to measure radial velocities and confirm the nature of GC candidates from \citetalias{Hughes2021}. We focus on radial velocity measurements using the strong
H$\beta$ (4861 \AA) and magnesium triplet (Mg b, 5167-5183 \AA) absorption features.

Owing to the relatively smaller field of view of M2FS compared to AAOmega, we generally observed fields closer to the central regions of NGC~5128 due to the higher density of good GC candidates at these projected radii. The exception is pointing 5, which was positioned to capture CenA-Dw1 and CenA-Dw3, two important stellar substructures discussed further in Section~\ref{sec:GCsub}.
Within each field we prioritized candidates based on preliminary versions of our \citetalias{Hughes2021} cluster catalog.
Across the eight M2FS fields, we measured spectra for 750 unique targets. 
Some of these targets were measured in multiple fields, and we also intentionally re-observed some velocity-confirmed GCs to test the fidelity of our velocity measurements, as discussed in Section~\ref{sec:radial_velocities}.

Table \ref{table:m2fs_observing} lists the date, field center location in right ascension and declination in J2000 coordinates, and the total exposure times for each field. Figure \ref{fig:map_observed} illustrates the M2FS field center positions with respect to NGC~5128 and the PISCeS data set (along with our AAOmega fields). Typically, each field was observed over the course of one night, though Field 8 was split over two observing runs in 2019.  The data from the two observing runs were treated separately, and independent velocity measurements were extracted from each. The Field 7 exposure time is lower than for the other fields due to a technical issue that occurred with the telescope during observing.

The M2FS observations are made in queue mode, and observing blocks are typically scheduled in dark or gray time. All of the data was taken in relatively clear skies and $\sim$ 0.7--1.2\arcsec\ seeing. Individual exposures of typically 45 minutes were combined into the final spectra.  In addition to science images, we also took a set of calibration frames in the afternoon or during the night, including twilight flats, biases, darks, thorium argon (ThAr) wavelength calibration arcs, and fiber maps.

Data reduction was performed using standard IRAF \citep{1986SPIE..627..733T} routines, as discussed in \citet{Walker2015}. 
First, we apply an overscan correction and trimmed each raw image, and applied zero level and dark count corrections. 
Cosmic rays are also identified and interpolated over using L.A. Cosmic \citep{2001PASP..113.1420V}.
Individual science exposures are combined into a single, stacked science frame from which scattered light is removed.
Then we trace fiber positions using twilight flat images, and extract one-dimensional spectra. We perform wavelength calibration using ThArNe or ThAr 1D lamp spectra.  Lastly, the spectra are normalized and sky-subtracted. We used the resulting science spectra and associated variance spectra.

%%%%%%%%%%%%%%%%%%%%%%%%%%%%%%%%%%%%%%%%%%%%%%%%%%%%%%%%%%%%%%%%
\subsection{AAT/AAOmega}

We obtained additional multi-object fiber spectroscopy using the 2dF/AAOmega instrument on the AAT 3.9 m telescope at the Siding Springs Observatory in Australia in 2017, 2019, and 2022. 
2dF is a multi-object fiber-feed to the dual-arm AAOmega spectrograph that is designed to allow the acquisition of up to 400 simultaneous spectra of objects across a two degree field on the sky \citep{Sharp2006}.
For the blue arm, we used the 580V grating, corresponding to a resolution of $R$$\sim$1300, and a wavelength range of $\approx$~3500--5500 \AA\ to cover the H$\beta$ and Mg b lines.  
For the red arm, we used the 1000I grating, corresponding to a resolution of $R$$\sim$4400, and a wavelength range of $\approx$~8200--9300 \AA\ to cover the calcium triplet (Ca) lines.  Our observing program was designed with the red arm in mind, but we use the blue arm spectra as a useful cross-check. The projected fiber diameters vary between 2.0--2.1\arcsec\ across the field of view.

Due to its very wide field, we used 2dF/AAOmega to cover the outer regions of NGC~5128.
Table \ref{table:2df_observing} lists the date, field center location in right ascension and declination in J2000 coordinates, and the total exposure times of our pointings. Figure \ref{fig:map_observed} shows the relative positions for each field. Seeing ranged from 1.4\arcsec--4.0\arcsec\ on the different observing nights, and we made adjustments to the exposures times to partially account for these variations. All individual exposures were 30 min.

We prioritized GC candidates with higher \textit{total\_likelihood} values from our selection process, although in 2017 and 2019 the selection methodology of \citetalias{Hughes2021} was not yet finalized.  In addition to science images, we also took arcs, flats, and bias images.  Across the six AAT fields, we measured velocities for 1780 unique targets.
Similar to our procedure with M2FS, we intentionally observed some GC candidates in multiple fields and re-observed confirmed GCs to test the fidelity of our velocity measurements.

We used {\sc 2dfdr}, the data reduction package for AAOmega, to reduce our spectroscopic data\footnote{\url{https://aat.anu.edu.au/science/software/2dfdr}}.  We extracted light from the fibers with optimal extraction and used a third order polynomial for the wavelength solution. For the red arm, the relative intensities of the skylines in the object data frames were used to determine the relative fiber throughput, using the {\sc Skyline(KGB)} algorithm within {\sc 2dfdr}.  Because the wavelength range of the blue arm had few bright skylines, we instead adopted the 
fiber-to-fiber normalization values from the red arm. The final product of {\sc 2dfdr} used in this paper was the combined spectrum of each target and an associated variance array.

%%%%%%%%%%%%%%%%%%%%%%%%%%%%%%%%%%%%%%%%%%%%%%%%%%%%%%%%%%%%%%%%%%%%%%%%%%%%%%%%%%%%
%%%%%%%%%%%%%%%%%%%%%%%%%%%%%%%%%%%%%%%%%%%%%%%%%%%%%%%%%%%%%%%%%%%%%%%%%%%%%%%%%%%%
\section{Radial velocities}\label{sec:radial_velocities}

In this section, we discuss our methodology for measuring radial velocities for our spectroscopic sample.  Following this, we compare our measurements with existing GC radial velocity measurements in the literature, and present our sample of newly confirmed GCs in NGC~5128.  With these results in hand, we evaluate our GC candidate selection technique reported in \citetalias{Hughes2021}.

\subsection{Measuring radial velocities}

We use similar procedures for measuring radial velocities in both our M2FS and AAOmega data sets, employing a cross-correlation procedure over different wavelength ranges.  We ensure robust velocities by cross-checking our results between these different wavelength ranges, along with visual inspection of individual velocity measurements.

\begin{figure}[t]
\centering
\includegraphics[width=1.0\linewidth]{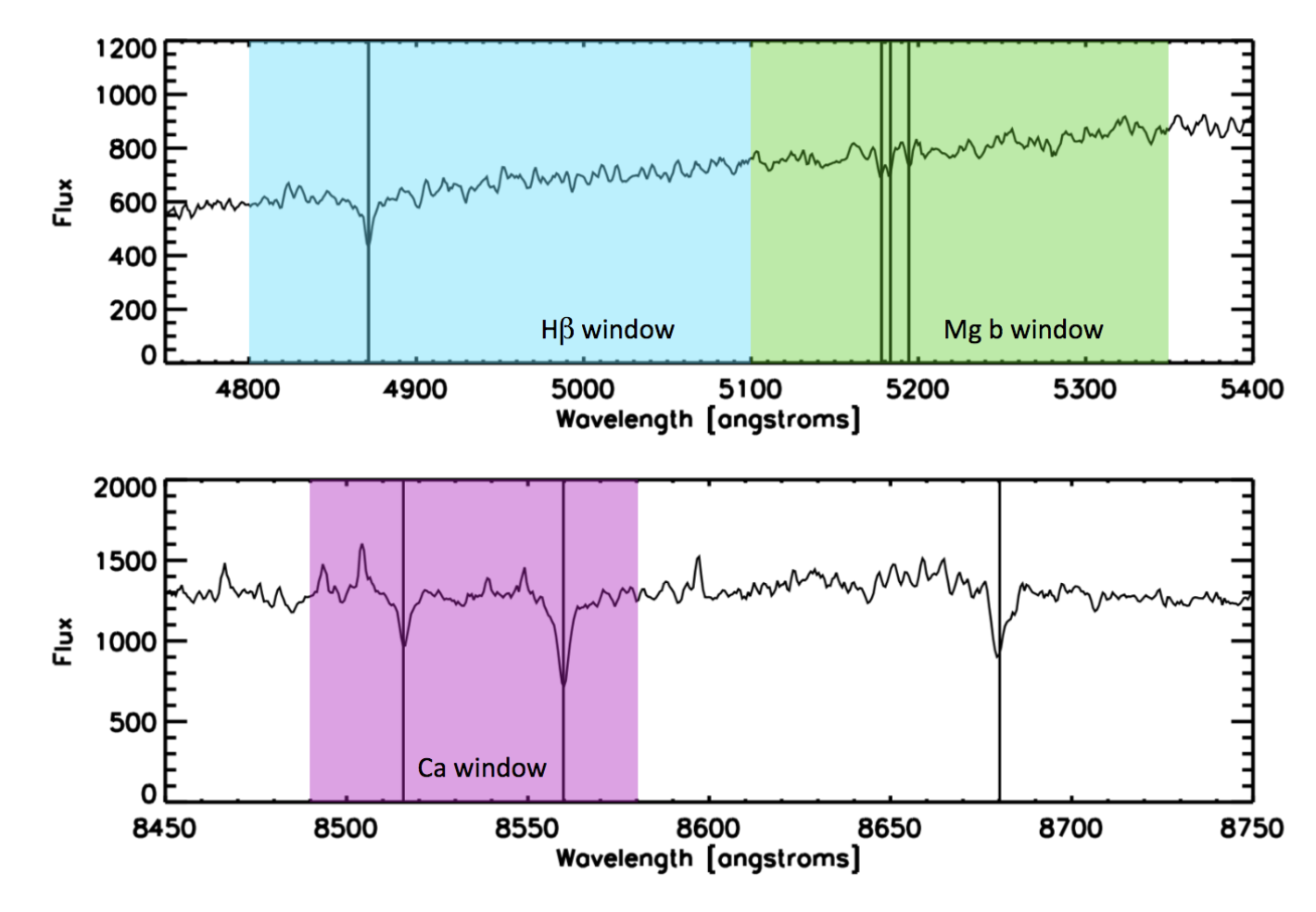}

\caption{Wavelength windows used to compute radial velocities via Fourier cross correlation in \texttt{fxcor} in IRAF.  In this example, we measure the target to have a radial velocity of 620 km s$^{-1}$. For AAOmega, all displayed regions are used; for the M2FS observations, only the H$\beta$ and Mg b triplet windows are used due to our wavelength coverage. 
}
\label{fig:windows}
\end{figure}

We compute radial velocities for our spectra via Fourier cross correlation, using  \texttt{fxcor} in IRAF. For the bluer data (M2FS and blue AAOmega data), we use an extremely high signal-to-noise (S/N) composite spectrum of M31 GCs as a template. We cross-correlate this around the H$\beta$ and Mg b spectral lines over the wavelength windows shown in Fig.~\ref{fig:windows}, noting that the latter region also includes Fe lines that can also be strong in metal-rich GCs. We compared the results using this composite template both to those obtained by using  candidate template stars of various spectral types obtained with the same instrument and also to synthetic model spectra, and found that the M31 composite best reproduced existing high-precision velocities with fidelity. Lacking a comparable template in the red Ca triplet region, we instead used a template based on the twilight solar spectrum. All velocities were corrected to the barycentric frame.

A well-known challenge to determining radial velocities via cross-correlation is the possibility of a ``catastrophic" misidentification of the  principal cross-correlation peak, which is a particular problem for low S/N spectra, especially when the noise is non-Gaussian as can arise from poorly subtracted sky lines. To address this potential issue, we require consistent cross-correlation results between at least two spectral features, and we also visually checked the candidate velocities for all spectra. 
In total, we measure accurate velocities for a total of 964 objects.  
For the M2FS data that resulted in a velocity measurement,  we calculated a median S/N (per pixel) of 12.9 (14.5) and $R$ value of 20.5 (26.9) for the H$\beta$ (Mg b triplet) feature, while for the AAOmega data we have a median S/N (per pixel) of (10.3, 10.5, 17.7) and $R$ value of (19.6, 23.3, 28.6) for the H$\beta$, Mg b triplet and calcium triplet, respectively.  Here the $R$ value is a measure of the cross-correlation peak height with respect to the rest of the cross-correlation function \citep[][]{Tonry79}, and is a standard output of the \texttt{fxcor} routine.

Regarding uncertainties, for the M2FS data we have a sufficient number of repeat observations between different fields to allow an estimate of the velocity uncertainties as a function of S/N. Following the procedure in \citet{2015AJ....149...53K}, we calculated the velocity differences between the pairs of spectra that both passed our quality criteria. We then binned these pairs in S/N, and calculated the spread of the paired velocity differences within each bin (using the median absolute deviation as a robust measure of the spread). Finally, we fit a relation between the S/N and this measure of the spread. We use this relation to assign an uncertainty to each velocity based on the S/N of the parent spectrum, where the uncertainty is related to this spread by a factor close to unity (see \citealt{2015AJ....149...53K} for additional discussion). On the basis of existing GC radial velocity studies at similarly low spectral resolution and comparable S/N (e.g., \citealt{Strader2011}), we set a minimum radial velocity uncertainty of 10 km s$^{-1}$ for the M2FS and AAOmega velocities.

Because we did not have many targets that were observed in more than one AAOmega field, we estimated uncertainties using a similar methodology, but applied instead to the velocities measured separately from the blue and red arm spectra.
We fit a power law to the binned median difference in radial velocity measurements between the blue and red arms as a function of S/N. 
The error assigned to a given spectrum is based on its S/N and the power-law fit to these median values, with a floor uncertainty of 10~km~s$^{-1}$ for all of our measurements.

The final reported M2FS and AAOmega heliocentric radial velocity measurements are weighted averages of the respective velocity measurements (see the windows in Figure~\ref{fig:windows}) that passed visual confirmation. Our radial velocity values for previously confirmed GCs in the literature are consistent within our measured uncertainties, as discussed in the next section.

\subsection{Literature comparison} \label{sec:lit_comp}

\begin{figure*}[th!]
\centering
\includegraphics[scale=0.4]{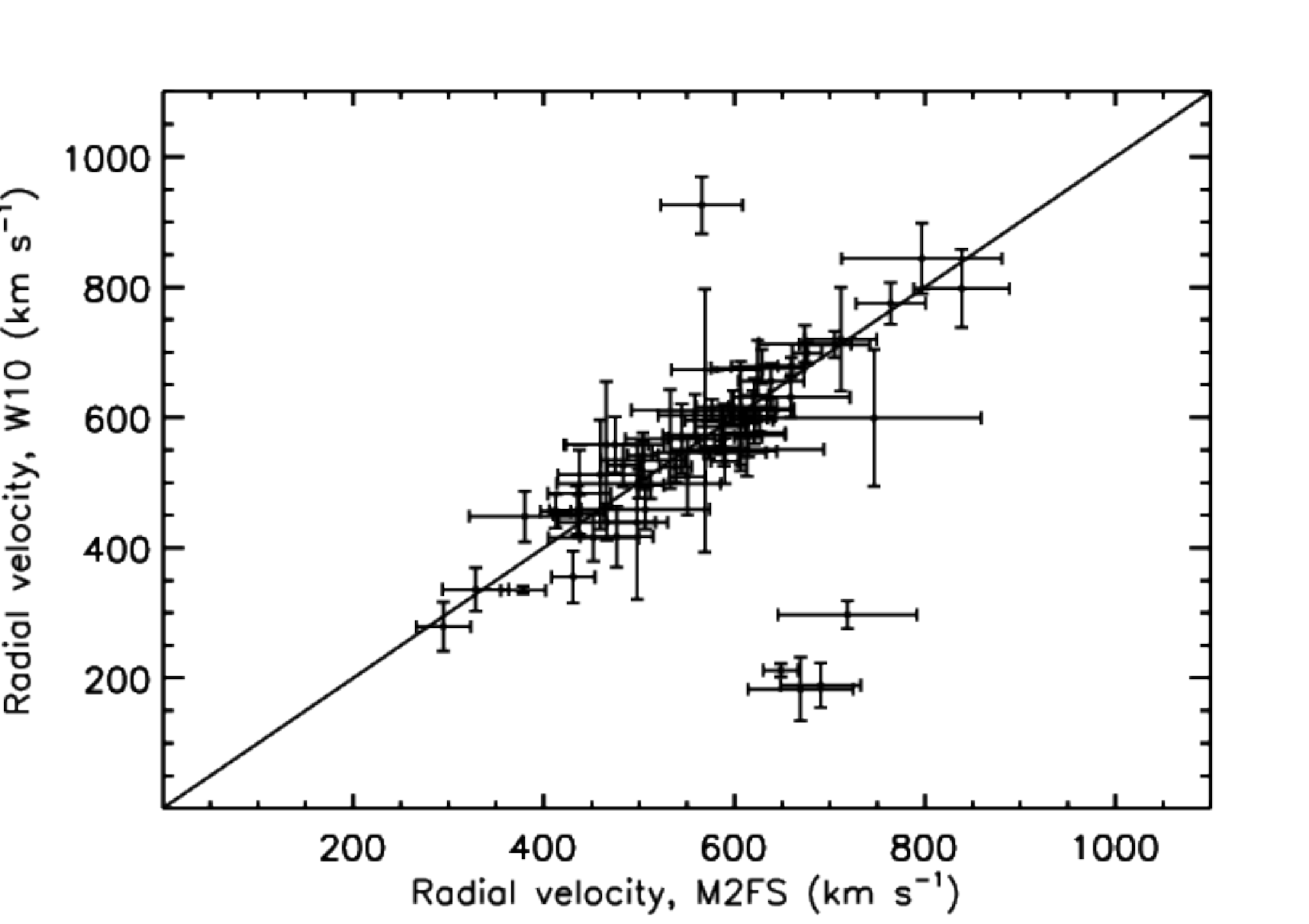}
\includegraphics[scale=0.4]{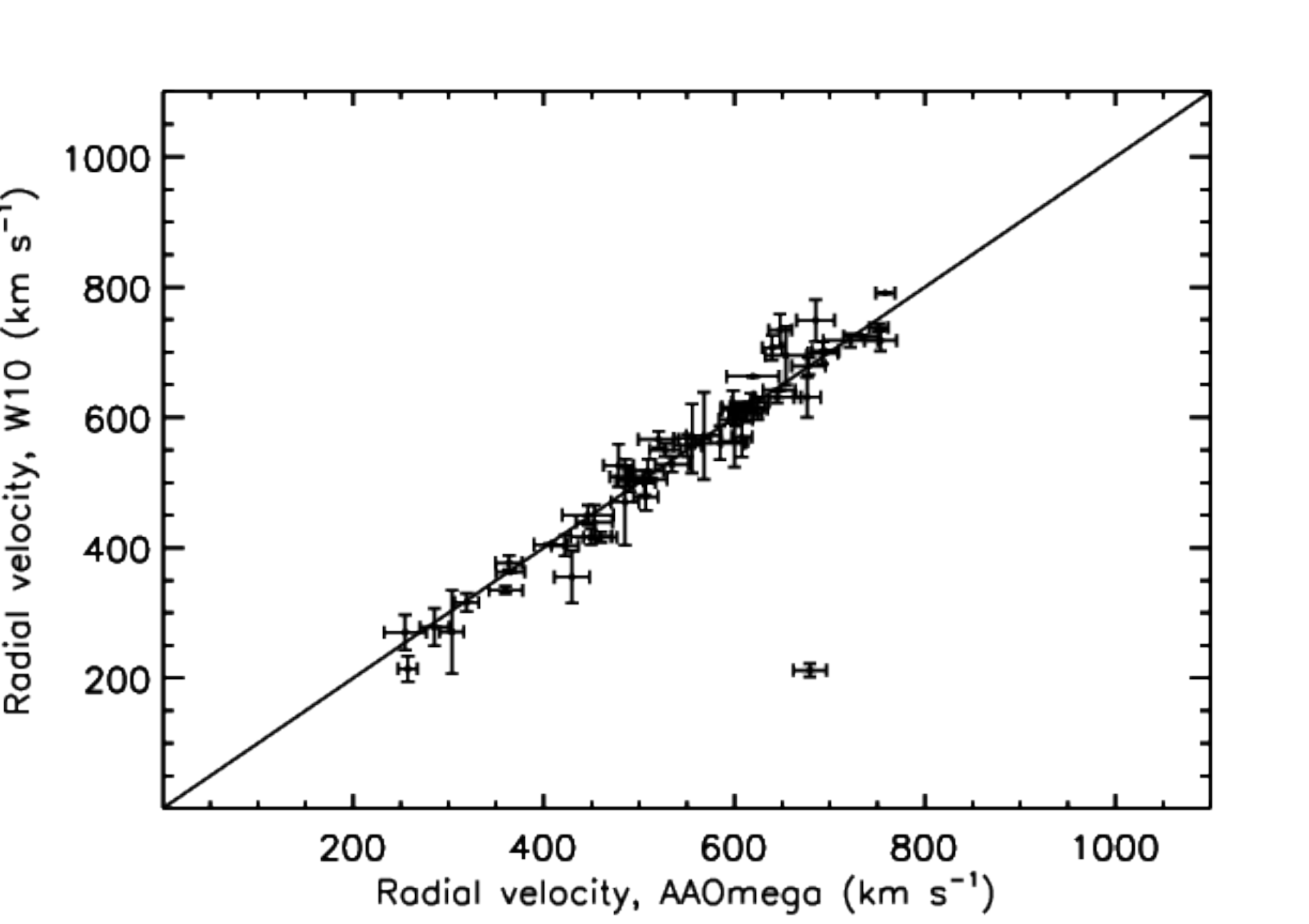}

\caption{We compare radial velocities measured from M2FS (left) and AAOmega (right) to the weighted radial velocity measurements in the \citetalias{Woodley2010a} catalogs, showing that with a small number of exceptions our velocities agree with published velocities with high fidelity. A 1:1 line is shown.
The outlying, inconsistent sources are discussed in the Appendix.}
\label{fig:vel_comp}
\end{figure*}

During our observations with M2FS and AAOmega we intentionally observed previously confirmed GCs to test the fidelity of our velocity measurements and to reduce uncertainties associated with velocities of confirmed GCs.  
We use the catalog of confirmed GCs from \citet{Woodley2010a}, hereafter \citetalias{Woodley2010a}, as reference because they are the most recent survey to publish a large sample of composite velocity measurements.
We obtained new radial velocities for 108 confirmed GCs %v7
from the \citetalias{Woodley2010a} catalog.  In Figure~\ref{fig:vel_comp} we show a comparison of our measured velocities with the weighted velocities from \citetalias{Woodley2010a}. 
Overall, the measurements agree very well, with a median difference of $-$3$\pm$3 km s$^{-1}$ (1$\pm$7 km s$^{-1}$) between the AAT (M2FS) data and \citetalias{Woodley2010a}, where the uncertainty is a robust rank-based estimate of the standard deviation divided by the square root of the number of measurements.

While \citetalias{Woodley2010a} was the largest single previous compilation of velocities, other subsequent studies contributed additional data, and overall there are 174 previously confirmed GCs with radial velocities that we re-observed. Table \ref{table:known_gcs} lists the GC ID from \citetalias{Hughes2021} or the present work, the right ascension and declination in J2000 coordinates, the PISCeS $g$ and $r$ magnitudes, the PISCeS $(g-r)_0$ and NSC $(u-z)_0$ color with Milky Way dust correction applied on a source-by-source basis \citep{Schlafly2011}, and the new velocity measurements from M2FS and AAOmega.  It also lists the \textit{new} weighted velocity measurement combining new measurements from the current work and all previous velocity measurements used in the weighting, as listed in the final column. The astrometry listed is primarily from PISCeS (see \citetalias{Hughes2021}), but not in all cases, since some sources are outside of the usable PISCeS area or are saturated. Hence, the astrometry listed is not homogeneous, but should be of sufficient precision to unambiguously identify all sources.

Among these repeats, there are only a few cases of greatly differing velocity measurements between the current survey and literature (see Figure~\ref{fig:vel_comp}). We discuss each of these in detail in the Appendix, along with a handful of conflicting measurements in the literature which we were able to resolve with our new spectroscopic data.

Because of the high rate of foreground star contamination among GC candidates with velocities $< 250$ km s$^{-1}$ (see discussion below), we only include GCs in our final catalog of confirmed objects if their radial velocities are $> 250$ km s$^{-1}$ or if they are clearly resolved in high-resolution imaging from the Hubble Space Telescope.

\subsection{Newly confirmed globular clusters}
\label{subsec:confirmed_clusters}
NGC~5128 has a systematic velocity of 541 km s$^{-1}$  \citep{Hui1995}, and a central stellar velocity dispersion of $\sim150$ km s$^{-1}$ \citep{Wilkinson1986}.  Therefore, to avoid contamination from Milky Way foreground stars, past surveys have generally considered a GC to be confirmed if its radial velocity is $\gtrsim 250$ km s$^{-1}$, at about $2\sigma$ below systemic. Some studies have confirmed  GCs at lower velocities, or without a velocity measurement at all, based instead on structural parameters (for example, being resolved in ground-based or space-based imaging). 

For our new data, we classify candidates as GCs if they have radial velocities greater than 250 km s$^{-1}$ and lack contravening evidence (i.e. a significant proper motion). Below this limit, we considered objects as potential GCs on a case-by-case basis, though we ultimately did not find indisputable evidence for additional GCs in this velocity range in our new data. This is not too surprising given that perhaps $\sim 2\%$ of our sample might be expected to have such low velocities. Our ability to separate foreground stars from low-velocity GCs with confidence is substantially improved compared to most previous work on NGC 5128 given the availability of Gaia, which allows the rejection of foreground stars as objects with significant proper motions that might otherwise be classified as potential GCs.

We find a total of 122 new NGC~5128 GCs based on our radial velocity measurements.  Table \ref{table:new_gcs} lists the ID from \citetalias{Hughes2021} or here, 
the right ascension and declination in  J2000 coordinates (with the same caveats on the astrometry as for Table \ref{table:known_gcs}), the PISCeS $g$ and $r$ magnitudes not corrected for dust, the PISCeS $(g-r)_0$ and NSC $(u-z)_0$ color with Milky Way dust correction applied on a source-by-source basis \citep{Schlafly2011}, and the measured radial velocities from M2FS and AAOmega.

\begin{figure*}[th!]  
\centering
\includegraphics[width=0.9\linewidth, trim={0 0 2.25cm 0} ,clip] {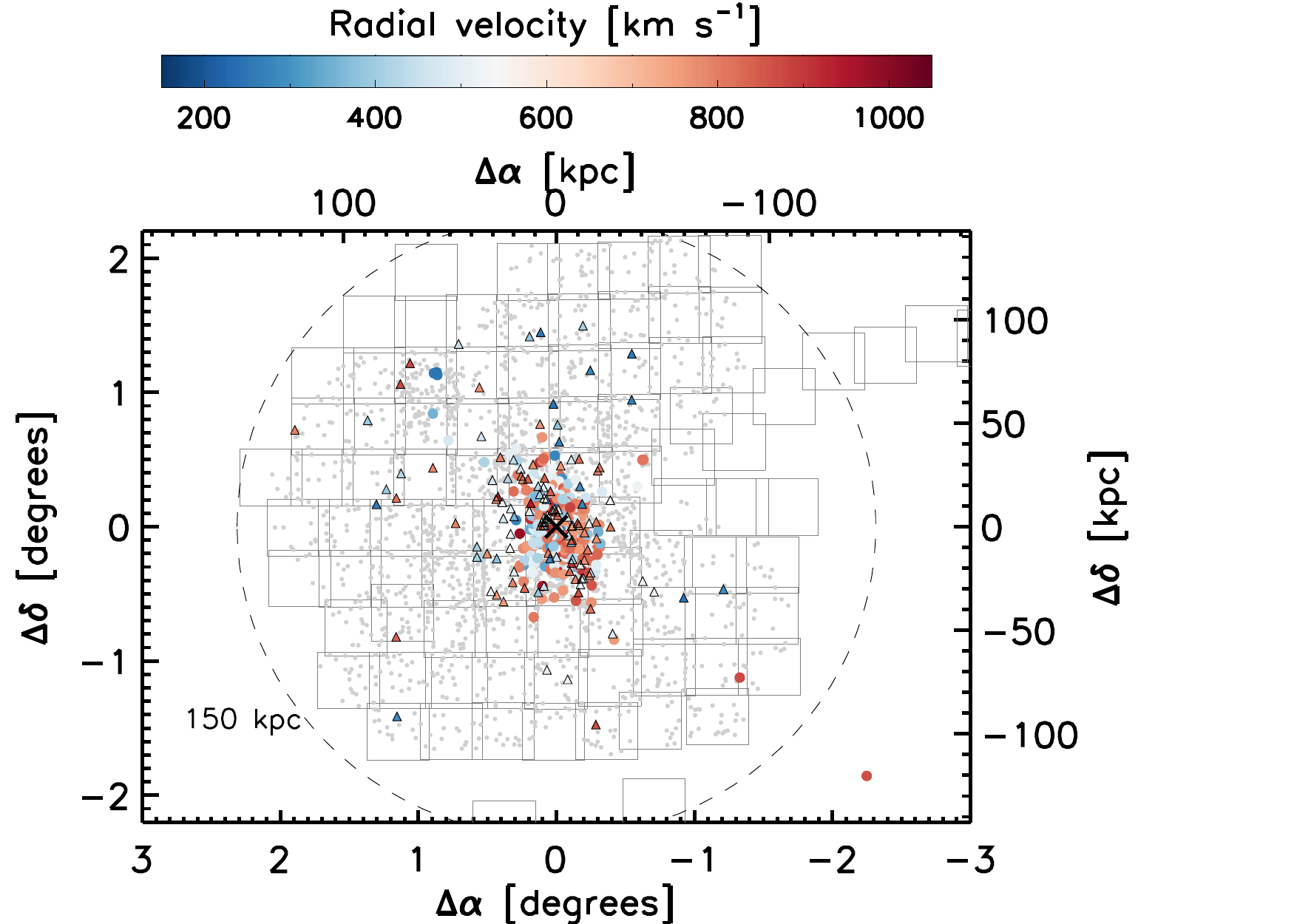}
\caption{Positions of all confirmed GCs in NGC~5128 with radial velocity measurements, where symbol color corresponds to radial velocity.  Triangles mark newly identified GCs, and circles mark previously confirmed GCs. To demonstrate our observing coverage, grey points mark positions of spectroscopically observed targets that were not confirmed to be GCs. An `X' marks the center of NGC~5128, with a velocity of 541 km s$^{-1}$.
}
\label{fig:map_velocity}
\end{figure*}

As a final check for both our new GC candidates and for those with only radial velocity measurements from the literature, we cross-match all candidate GCs against Gaia DR3 \citep{DR3} to see if they have measured proper motions or parallaxes that could suggest a foreground star classification (recall that our GC selection technique used for spectroscopic targeting employed Gaia DR2). Any candidates with high significance ($5\sigma$) Gaia DR3 proper motions and radial velocities $< 350$ km s$^{-1}$ are classified as stars. For a few objects the evidence is mixed: these cases are discussed in more detail in the Appendix.
Not all targets have Gaia data, so this evaluation is partially incomplete, but it still represents a meaningful advance in the purity of the final GC sample.

We plot the positions of all confirmed GCs with radial velocities in Figure \ref{fig:map_velocity}. We discovered 28 new GCs at large radii, at $>$50\arcmin~($\approx$54 kpc), with the most distant new object at a radius of 122\arcmin~($\approx$130 kpc). These newly confirmed distant GCs are primarily to the north of the galaxy, though we emphasize that this region also benefited from the finalized GC candidate selection technique and received the most amount of observing time. An additional 69 new GCs are within 30\arcmin~ ($\approx$32 kpc), which illustrates that substantial scope exists for confirmation of more centrally located candidates as well.

As a resource for future follow-up observations, we also include a table of 656 stars and 12 galaxies identified via radial velocity measurements from M2FS and AAOmega in Table~\ref{table:bad}.

\subsection{Evaluation of GC selection technique}\label{sec:roi}

In \citetalias{Hughes2021}, we highlighted GC candidates of gold and silver rank with \textit{total\_likelihood} $\geq~0.85$ as the most likely to be true GCs in NGC 5128 and therefore the most promising targets for follow-up spectroscopic confirmation. We refer to this combined sample here as the ``priority" sample. 
Because our spectroscopic observations were taken over a few years while we were still refining our techniques, our GC candidate selection process changed between observation runs. This means that the earlier spectroscopic observations were more likely to include contaminating foreground stars and background galaxies.

We observed 655 of the \citetalias{Hughes2021} priority sample targets during our observing runs with Magellan/M2FS and AAT/AAOmega.  Of these, 179 (27\%) are found to be new or previously confirmed GCs and 134 (20\%) are found to be foreground stars or background galaxies. The remainder could not be conclusively identified, typically due to low S/N. 
Of those GC candidates not in the priority sample, only 7\% are found to be new or previously confirmed GCs.

We expect the level of contamination to be higher in the outer regions of the galaxy where the density of GCs is lower. Dividing the sample at 30\arcmin~ (33 kpc), we find that within this radius, the fraction of confirmed GCs is 143 out of 209 (68\%), while beyond 30\arcmin, we confirm 36 out of 446 (8\%).

%%%%%%%%%%%%%%%%%%%%%%%%%%%%%%%%%%%%%%%%%%%%%%%%%%%%%%%%%%%%%%%%
%%%%%%%%%%%%%%%%%%%%%%%%%%%%%%%%%%%%%%%%%%%%%%%%%%%%%%%%%%%%%%%%
\section{Analysis of the updated GC population}\label{sec:analysis}

In this section we look at the two-dimensional and radial distributions of the updated confirmed GC population as a whole, as well as split into metal-poor and metal-rich subpopulations.  We identify GCs associated with stellar substructures seen in the RGB star map from the PISCeS survey.  Lastly, we look at the distribution of radial velocities and calculate an updated mean value and velocity dispersion for the GC system.

After our new measurement and analysis of literature objects in Section~\ref{subsec:confirmed_clusters}, the full sample of confirmed GCs with radial velocity measurements in NGC~5128 is now 645.  This number includes the 122 new measurements presented here. 
We plot the positions of these radial velocity confirmed GCs around NGC~5128 in Figure \ref{fig:map_velocity}.  While 88\% of the total confirmed GC population is within 30\arcmin, there are many GCs out to $\approx$100\arcmin~ ($\approx$108 kpc), with the farthest known GC at a projected radial distance of 173\arcmin~ (187 kpc).

%%%%%%%%%%%%%%%%%%%%%%%%%%%%%%%%%%%%%%%%%%%%%%%%%%%%%%%%%%%%%%%%
\subsection{Surface Density \& Metallicity trends}\label{sec:updated_color}

\begin{figure}[t]
\centering
\includegraphics[width=1.0\linewidth]{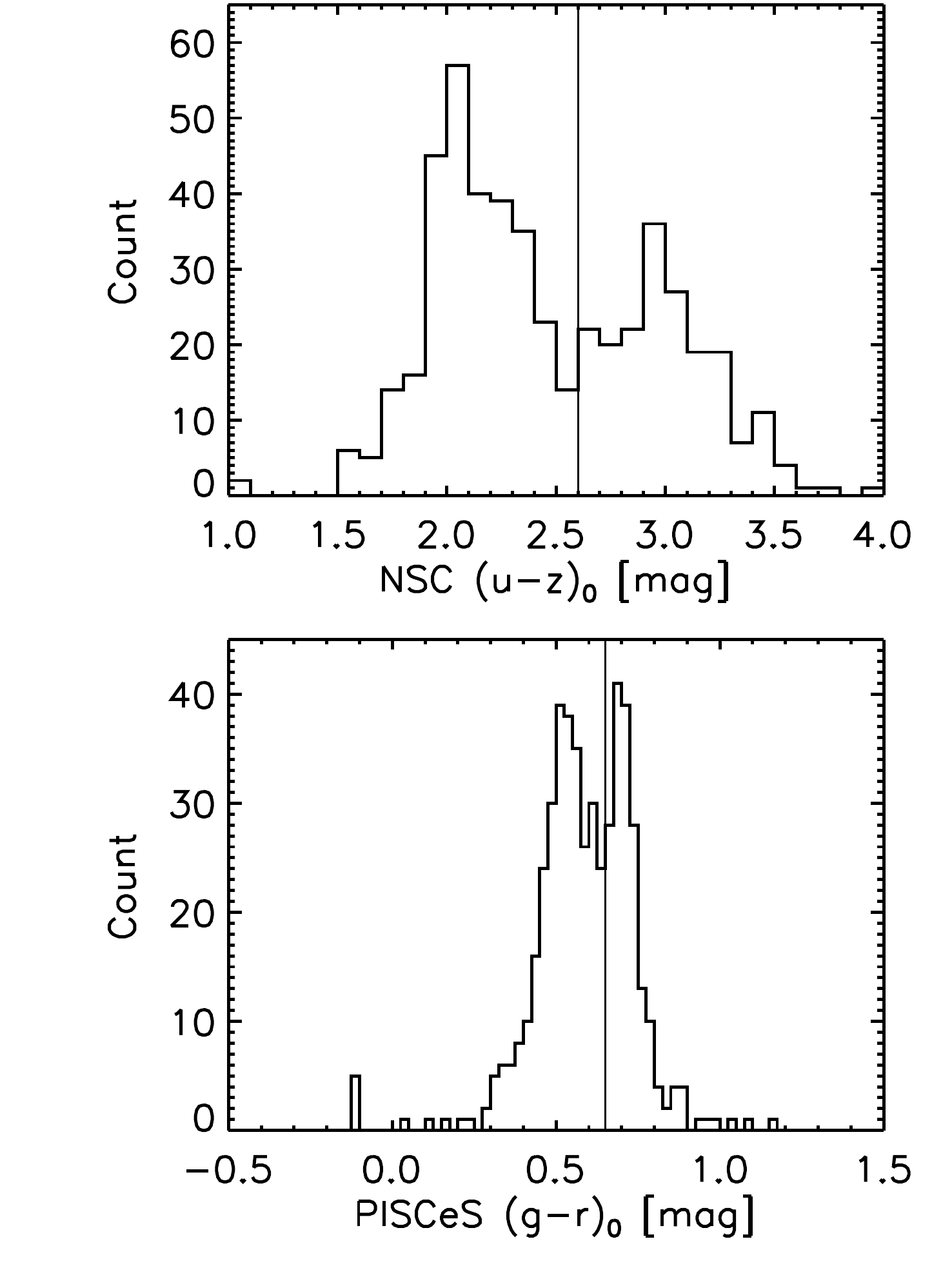}

\caption{Confirmed GCs in NGC~5128 show a bimodal color distribution, which we use to split them into metal-poor (blue) and metal-rich (red) subpopulations. The top histogram shows $(u-z)_0$ color from the NSC and bottom histogram shows $(g-r)_0$ color from PISCeS for the same set of GCs, where photometry for both has been corrected for foreground extinction.  The vertical lines show the adopted divide between the two subpopulations, at $(u-z)_0 = 2.6$ and  $(g-r)_0 = 0.65$ mag. 
}
\label{fig:color_decider}
\end{figure}

Given the clear observed bimodal color distribution of GCs in NGC 5128 (Figure \ref{fig:color_decider}), we make the usual subdivision of the GCs into either metal-poor (blue) or metal-rich (red) subpopulations based on their NSC colors.  We use the same divide as in \citetalias{Hughes2021},  such that GCs with $(u-z)_0 < 2.6$ are in the metal-poor sample and GCs with $(u-z)_0 > 2.6$ are in the metal-rich sample.  For those GCs that do not have photometry in NSC, we find an equivalent divide at $(g-r)_0 = 0.65$ based on PISCeS photometry, as shown in Figure \ref{fig:color_decider} (where GCs can appear in both the top and bottom panels). 
There are 9 GCs with radial velocity measurements that do not have photometry in either NSC or the PISCeS catalog, and as such they are not included in the subsequent color-based subpopulation analysis.  Most of these GCs without photometry are within 5\arcmin\ of the galaxy center, where ground-based photometry is difficult due to the extremely high and variable background of NGC~5128's central regions.
We emphasize that beyond our conservative overall color selection of GC candidates (see \citetalias{Hughes2021}), we did not preferentially follow-up blue or red GC candidates.

\begin{figure}[t]
\centering
\includegraphics[width=1.0\linewidth]{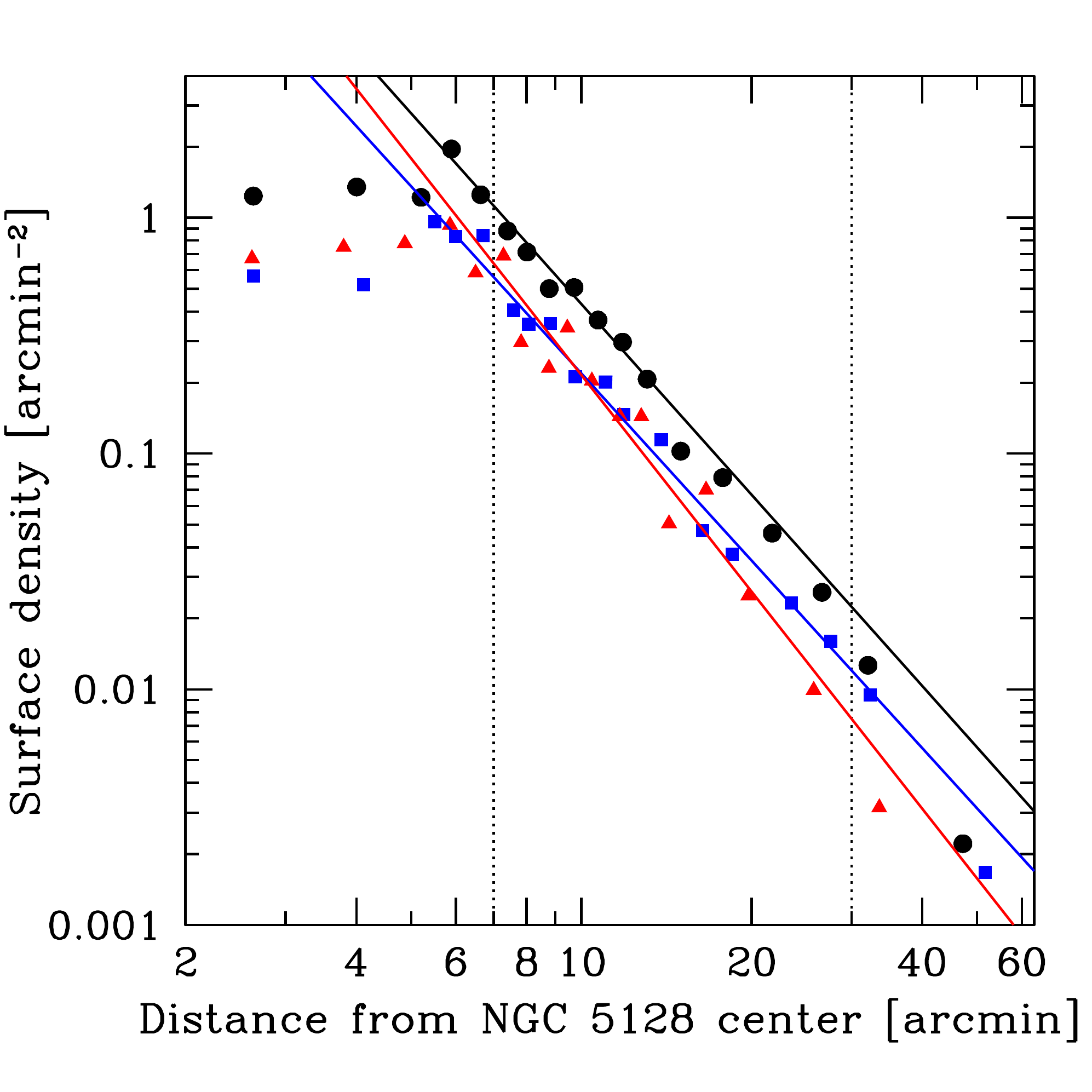}
\caption{Binned surface density profiles for velocity-confirmed GCs with $20.5 > r > 17.5$, showing the steeper radial profile of the metal-rich GCs than the metal-poor GCs. These surface densities are calculated in circular annuli with near-constant numbers of GCs. The samples are all GCs (black circles), metal-poor GCs (blue squares), and metal-rich GCs (red triangles). The overplotted power-law fits are made from radii 7\arcmin--30\arcmin; these fitting limits are denoted with dotted lines.
Within 7\arcmin, the surface density profiles flatten due to incompleteness.  As a reminder, 1 arcmin is equal to 1.1 kpc.
}
\label{fig:radial_color}
\end{figure}

We find that 55\% of the confirmed GCs are metal-poor and the other 45\% are metal-rich.  Within $\sim 10\arcmin$, there are approximately equal numbers of metal-poor and metal-rich GCs.  Beyond this, metal-poor GCs begin to dominate.  
The radial distributions of velocity-confirmed GCs with $20.5 > r > 17.5$ are plotted in Figure \ref{fig:radial_color}, where we have imposed the faint magnitude limit to ensure there is no bias between the M2FS and AAOmega samples, and the bright limit to guard against the possibility that stripped nuclei \citep{Dumont2021} could affect our results. These plotted surface densities are calculated in circular annuli with near-constant numbers of GCs.

As discussed above, the GC sampling is incomplete in the inner regions of the galaxy. We find that over a radial range of 7\arcmin--30\arcmin, power laws appear to provide excellent fits to the surface density of the total sample, as well as the blue and red subpopulations. The best-fit power-law indices for the total, blue, and red subpopulations are $-2.69\pm0.19$, $-2.64\pm0.27$, and $-3.05\pm0.28$, respectively, and these fits are plotted in Figure \ref{fig:radial_color}. The uncertainties in these fits are derived from bootstrapping (resampling with replacement).  These power-law values for the surface density profile are converted to physical 3D density profile indices for use in Section~\ref{sec:total_mass} and our mass estimates.

The metal-rich subpopulation has a slightly steeper slope and is more centrally concentrated than the metal-poor subpopulation, which follows trends observed in other extragalactic GCs studies (e.g., \citealt{Brodie06, Faifer2011, Forbes2012}).

At the largest radii covered by these data, the spatial distribution of GCs is no longer smooth. As discussed further in Section~\ref{sec:GCsub}, GCs associated with stellar substructures in the outer halo of NGC~5128 make up a substantial fraction (at least a third, and possibly more) of the GCs beyond $\sim 50\arcmin$. The vast majority of the GCs associated with visible substructures are metal-poor,
consistent with the ongoing assembly of the metal-poor outer halo of NGC 5128.

%---------------------------------------------------------
\subsection{Radial velocity distributions}\label{sec:radvel}

\begin{figure}[t]
\centering
\includegraphics[width=1.0\linewidth]{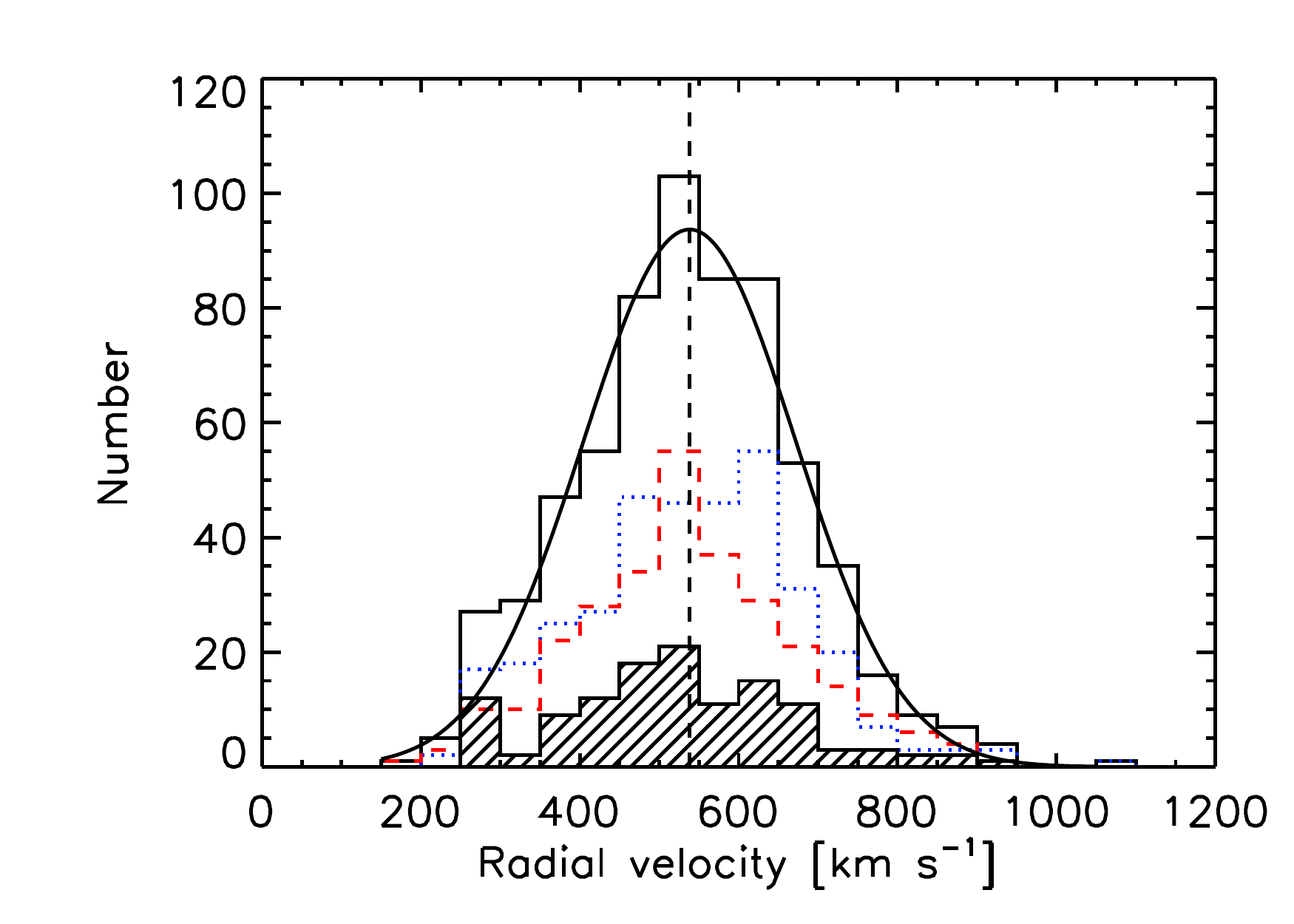}
\caption{Radial velocity distributions of the full confirmed GC population (tall histogram), newly confirmed GCs (hatched histogram), and metal-poor (blue dotted) and metal-rich (red dashed) subpopulations. We also overplot the Gaussian fit to the entire data set (see Section~\ref{sec:radvel}).
}
\label{fig:vel_hist}
\end{figure}

Radial velocity histograms are shown in Figure \ref{fig:vel_hist} for the entire confirmed GC system, as well as the subpopulations of metal-poor GCs, metal-rich GCs, and newly confirmed GCs from this study.
We fit a Gaussian function to the histogram of the entire GC population, and find a mean velocity of $538 \pm 5$ km~s$^{-1}$ and $\sigma = 134 \pm 4$ km~s$^{-1}$. 
The mean velocity of the GC system is consistent with the systematic velocity of NGC~5128 of 541 km~s$^{-1}$ \citep{Hui1995} with a velocity dispersion of $\sigma \sim 150$ km~s$^{-1}$ \citep{Wilkinson1986,Silge05}.  Consistent results are found for the metal rich ($v_{sys}$=535$\pm$9 km~s$^{-1}$; $\sigma$=139$\pm$7 km~s$^{-1}$) and metal poor ($v_{sys}$=538$\pm$7 km~s$^{-1}$; $\sigma$=131$\pm$5 km~s$^{-1}$) GC populations when considered separately as well.
Many of the GCs associated with the substructures marked in Figure \ref{fig:marked_map} and discussed in Section~\ref{sec:GCsub} have relatively low velocities, and excluding these GCs brings the mean velocity to $540 \pm 5$ km~s$^{-1}$, consistent with the sample as a whole.

%---------------------------------------------------------

\begin{figure*}[t]
\centering
\includegraphics[width=0.48\linewidth]{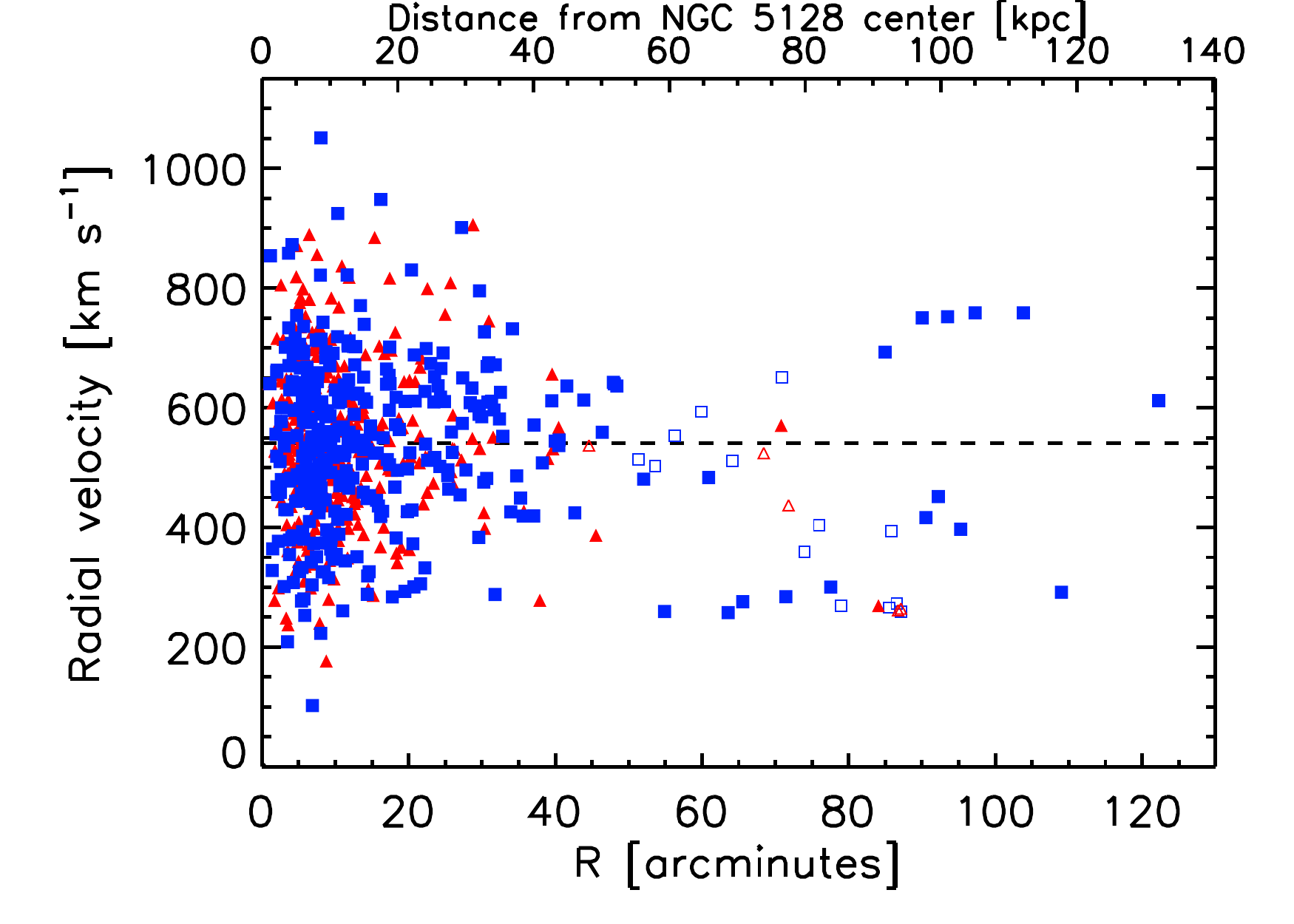}
~
\includegraphics[width=0.48\linewidth]{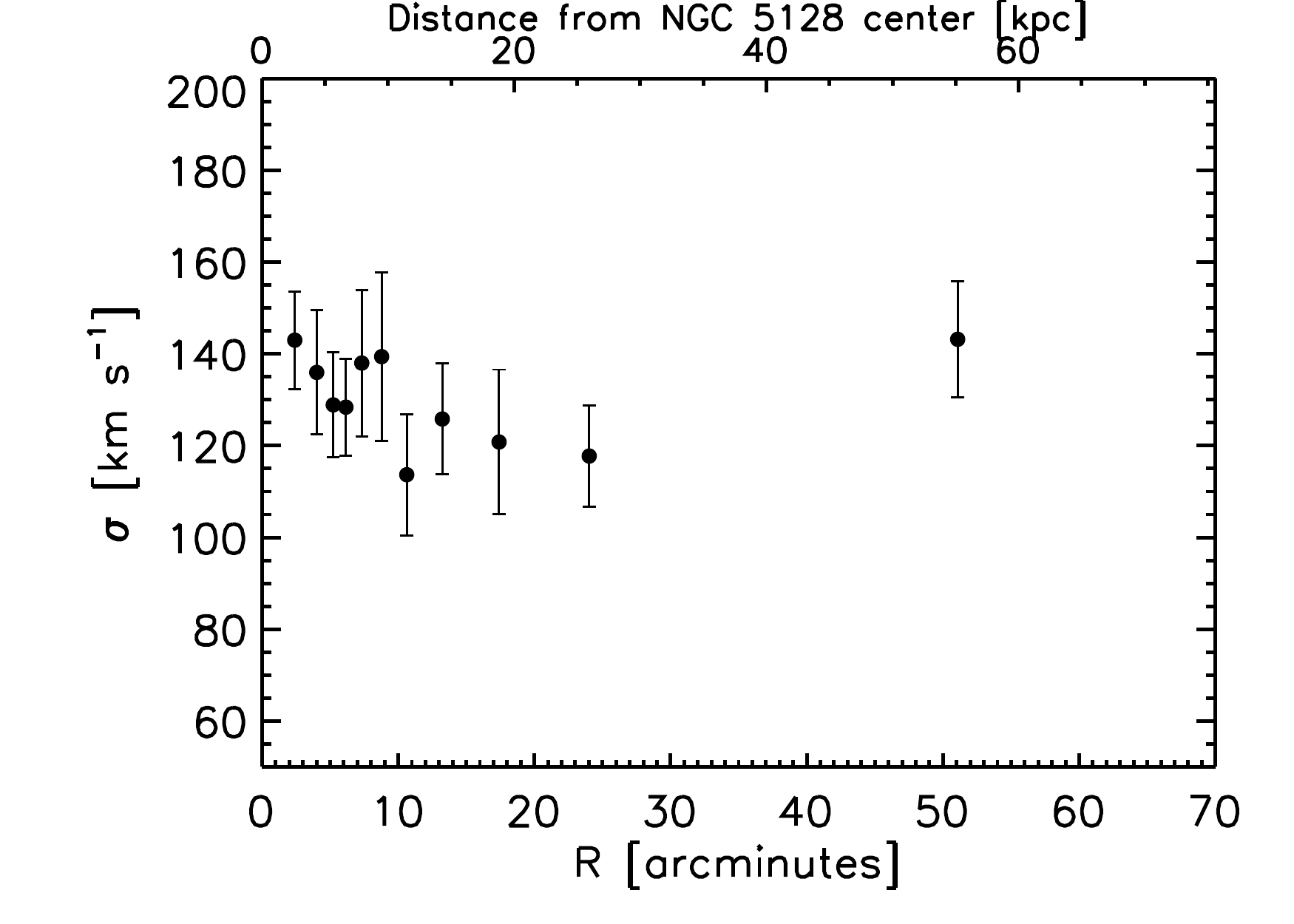}
\caption{Left: Radial velocity measurements for confirmed metal-rich (red triangles) and metal-poor (blue squares) GCs as a function of projected galocentric radius, where unfilled symbols denote GCs in the substructures marked in Figure \ref{fig:marked_map} and filled symbols are all other GCs.  The horizontal dashed line marks the systematic velocity of 541 km~s$^{-1}$ for NGC~5128. 
Right: Velocity dispersion as a function of radius, binned by groups of 57 GCs, using a maximum likelihood estimator \citep[e.g.][]{Pryor93}. 
}
\label{fig:vel_dist}
\end{figure*}

We show radial velocity and velocity dispersion measurements as a function of projected radial distance in Figure \ref{fig:vel_dist}.  
\cite{Woodley2010a} found that the metal-poor GCs in NGC~5128 out to 45\arcmin~ have lower radial velocities in the mean than the metal-rich GCs. In our larger sample we see no difference in the mean velocity between blue and red GCs, either for the whole sample or restricting it to radii $< 45\arcmin$. As seen in the right panel of Figure \ref{fig:vel_dist}, the velocity dispersion declines slowly from $\sim 140$ km s$^{-1}$ in the central regions to  $\sim 120$ km s$^{-1}$ at 25\arcmin. At large radii the velocity dispersion goes back up, likely due the presence of unrelaxed substructure (Section \ref{sec:GCsub}).

\begin{figure*}[t]
\centering
\includegraphics[width=1.0\linewidth]{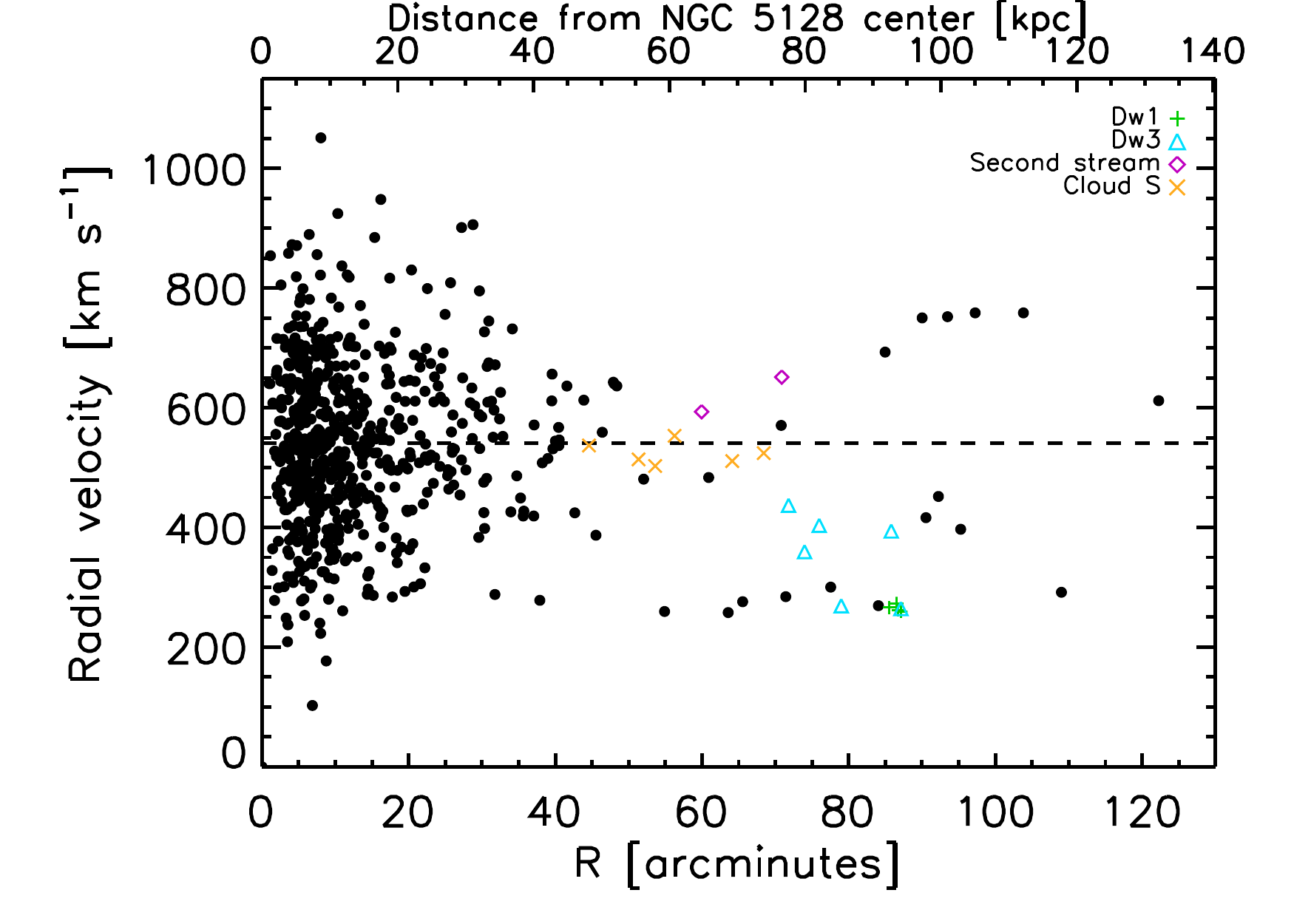}
\caption{Radial velocity measurements for confirmed GCs as a function of projected galactocentric radius, where colored symbols denote GCs associated with the substructures marked in Figure \ref{fig:marked_map} and black dots are all other GCs.  The horizontal dashed line marks the systematic velocity of 541 km~s$^{-1}$ for NGC~5128. 
}
\label{fig:vel_dist_substruc}
\end{figure*}

In Figure~\ref{fig:vel_dist_substruc}, we again show radial velocity as a function of projected radial distance, this time with symbols denoting the GCs associated with specific halo substructures from Figure \ref{fig:marked_map}. We see that the GCs that are spatially co-located with individual substructures also tend to cluster in radial velocity, indicating that at least in some cases the GCs and underlying stars are indeed associated coherent structures. 

By contrast, other GCs at large radii (80--110\arcmin), even those that clump together at velocities around $\sim 750$ km s$^{-1}$ (Figure \ref{fig:vel_dist_substruc}), are far from each other in projected two-dimensional space and do not appear to be obviously causally connected. It is possible for spatially distant objects to be related in phase space (for example, if they are members of a shell; \citealt{1998MNRAS.297.1292M}). Small number statistics leading to the false appearance of clustering may also be at play.

%%%%%%%%%%%%%%%%%%%%%%%%%%%%%%%%%%%%%%%%%%%%%%%%%%%%%%%%%%%%%%%%%%%%%%%%%%%%%%%%
% Analysis
%%%%%%%%%%%%%%%%%%%%%%%%%%%%%%%%%%%%%%%%%%%%%%%%%%%%%%%%%%%%%%%%%%%%%%%%%%%%%%%%

\section{Globular Clusters and Outer Halo Substructures}\label{sec:GCsub}

\begin{figure*}[t]
\centering
\includegraphics[width=0.8\linewidth]{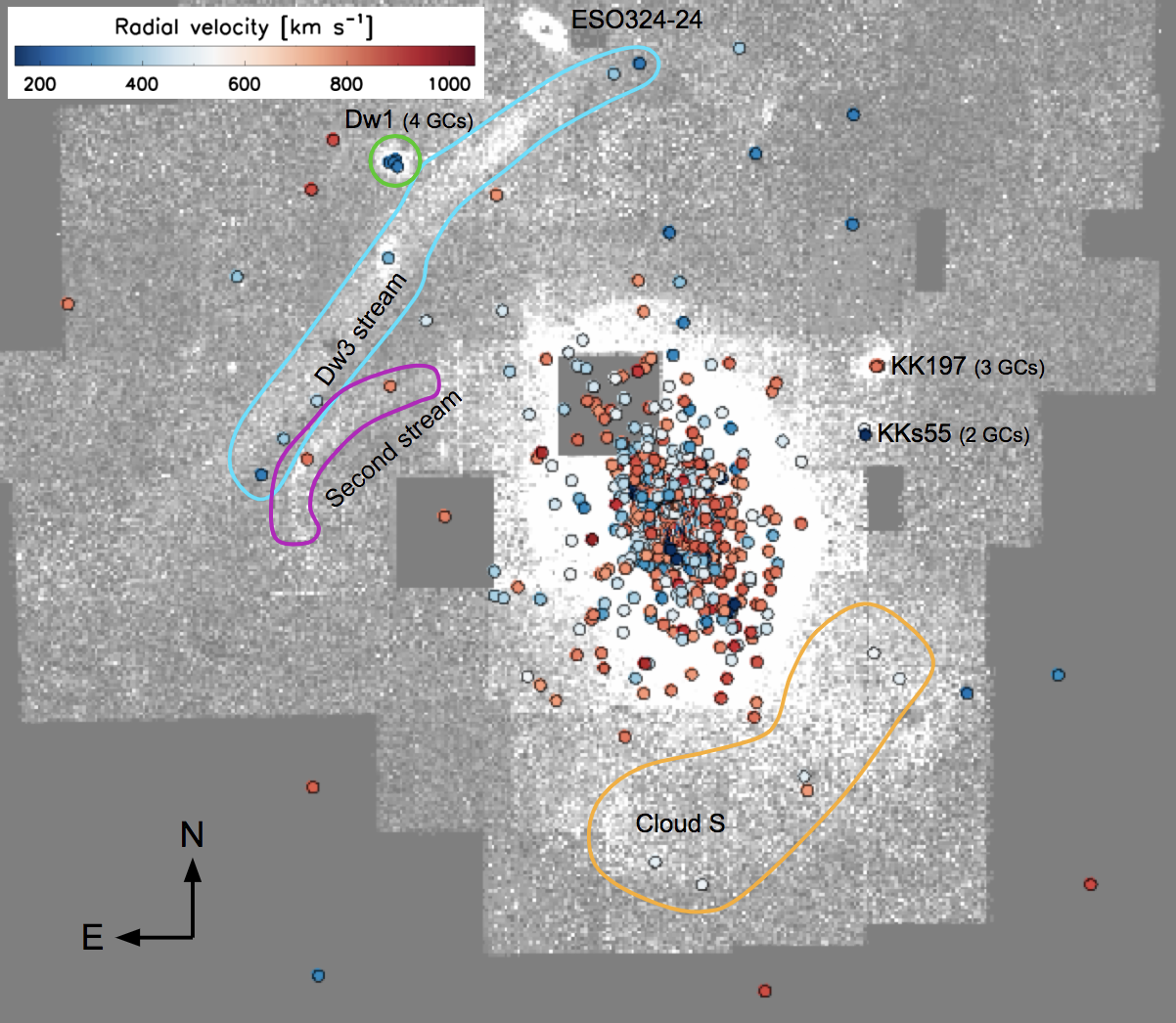}
\caption{RGB map of NGC~5128 from PISCeS.  Positions of known GCs are marked with colored dots that correspond to their weighted radial velocity measurements. Colored contours mark cold stellar features of interest discussed in the text. We refer the reader to Figure~\ref{fig:map_velocity} and \ref{fig:vel_dist_substruc} for further context. 
}
\label{fig:marked_map}
\end{figure*}

The outer halo of NGC~5128 is rich with various field substructures in the form of stellar streams and clouds \citep{2013MNRAS.432..832C,Crnojevic16,Crnojevic19}. These substructures track the galaxy's active and ongoing merger history.  Tidal streams also act as probes of the gravitational potential of the galaxy, and their morphology and kinematics can be used to constrain NGC~5128's dark matter halo mass \citep{Pearson22}.
However, determining velocities for these low surface-brightness features is very challenging \citep{Toloba16}. The kinematics of both tidal structures and dwarfs can sometimes be more easily derived from their associated GCs.  Previous work has used GC radial velocities to both study the kinematics of substructure in the halo of M31  \citep{Veljanoski2014}, and to constrain the dynamical mass of dwarf galaxies \citep[e.g.][]{VanDokkum2016,Toloba18}.

\citet{Crnojevic16} published a detailed map of the density of red giant branch (RGB) stars that shows many substructures and dwarf galaxies in the outer halo of NGC~5128.  Figure \ref{fig:marked_map} shows that many of our outer GCs appear spatially associated in projection with prominent stellar features in this map.  Furthermore,  the GCs associated with these features tend to exhibit similar radial velocities. We note that the RGB map shown here includes PISCeS imaging data available at the time \citet{Crnojevic16} was published, and thus covers a more 
restricted area than our GC catalog. 

In the rest of this section, we consider several of these tentative GC groups and assess the significance of these observed velocity patterns.  Identification of the GCs associated with substructure also is important for our mass analysis in Section~\ref{sec:total_mass}.

\subsection{CenA-Dw1}

One of the first discoveries of the PISCeS survey was CenA-Dw1 ($M_V$=$-$13.8; $r_h$=1.8 kpc; R$_{\rm CenA, proj}$=93 kpc) and its possible companion CenA-Dw2 ($M_V$=$-$9.7; $r_h$=0.4 kpc; R$_{\rm CenA, proj}$=92 kpc), which are only separated by $\sim$3 kpc on the sky \citep{Crnojevic14,Crnojevic19}.  Inspection of follow-up Hubble Space Telescope imaging of the CenA-Dw1 field revealed four clear, semi-resolved globular clusters associated with CenA-Dw1.  All of these were GC candidates in the H21 sample, and were targeted for spectroscopic follow-up using the high resolution mode of M2FS as presented in \citet{Dumont2021}.  These were all confirmed to be GCs, and indeed they have velocities within $\approx$14 km s$^{-1}$ of each other, as would be expected for the GC system of a faint dwarf galaxy; see Figure~\ref{fig:vel_dist_substruc}.

The four CenA-Dw1 GCs are between $\approx$5--60\arcsec\ ($\approx$90--1140 pc) from the center of the dwarf, and have absolute magnitudes ranging between $M_V$=$-$7.0 and $-$8.7 mag. The brightest GC ($r_0 = 19.0$; $M_V$=$-$8.7 mag) is located closest to the dwarf center and could potentially be a nuclear star cluster, as these are often slightly offset from the galaxy center \citep[e.g.,][]{Georgiev2009}.  Taking the weighted mean velocity of all four GCs, we find $v=265\pm3$ km s$^{-1}$ ($262\pm3$ km s$^{-1}$ if the central cluster is removed). Within the precision of our measurements, the intrinsic velocity dispersion of this GC system is consistent with zero. If CenA-Dw1 had a mass-to-light ratio similar to the Fornax dwarf galaxy \citep{2007ApJ...667L..53W}, the velocity dispersion would be $\gtrsim 10$ km s$^{-1}$, which should have been detectable given our GC velocity uncertainties in most instances. This preliminary result is intriguing and should motivate future, more precise radial velocity measurements that would allow the determination of the dynamical mass of CenA-Dw1.

\subsection{CenA-MM-Dw3 and its Tidal Tails}\label{sec:tidal}

One of the most striking substructure features is CenA-MM-Dw3, a disrupting dwarf galaxy positioned to the north-east of the galaxy center with tidal tails spanning over 1.5$^\circ$ \citep{Crnojevic16}. A confirmed GC (H21-360500) sits at the center of the main remnant of the disrupting dwarf. This object has a velocity of $359\pm2$ km~s$^{-1}$ and is likely the dwarf's nuclear star cluster \citep{Dumont2021}, so we take this velocity as the best estimate of that for the dwarf.

Projected along the tidal tails are an additional five NGC~5128-confirmed GCs with measured radial velocities, which have velocities within $\sim 100$ km~s$^{-1}$ of the dwarf's systemic velocity. The spatial location of the GCs in the outer regions of the stream, and possible offset from the main body of the stream, would be consistent with a GC population that started as more extended than the stars in the dwarf, and which would have begun to be tidally stripped earlier than the bulk of the stars in the dwarf.

\begin{figure}[t]
\centering
\includegraphics[width=1.0\linewidth]{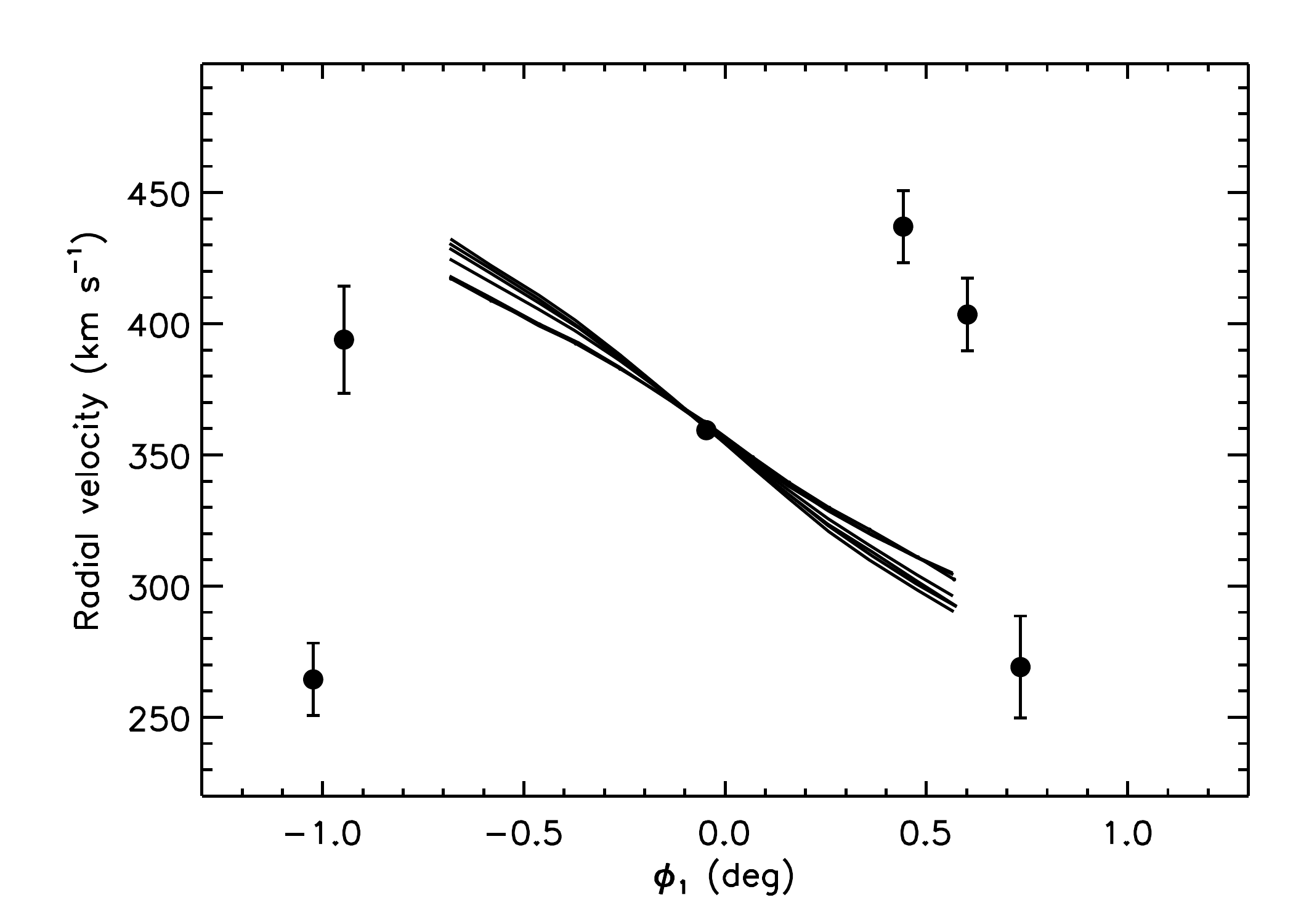}
\caption{Radial velocity as a function of distance from the position of the main galaxy remnant for the six GCs that lie projected on top of the CenA-MM-Dw3 stream. Over plotted are a series of dynamical models (evolved in potentials with different dark matter halo masses) for the predicted radial velocity gradient along the stream presented in \citet{Pearson22}, which both reproduce the stream morphology and the velocity of the central Dw3 star cluster (but were not fit against the new GC velocities presented here).  The x-axis coordinate ($\phi_1$) is adopted from \citet{Pearson22}, and represents the longitude in a coordinate system centered on NGC~5128 and rotated such that Dw3 is at $\phi_1$=0 degrees.  As we discuss in Section~\ref{sec:tidal}, it is not likely that all of the GCs projected onto the Dw3 stream are physically associated with it.  The sign of the velocity gradient would flip if CenA-MM-Dw3 were in the foreground with respect to NGC~5128, as described in Section~5.
}
\label{fig:dw3_gradient}
\end{figure}

Figure \ref{fig:dw3_gradient} plots the velocities of these GCs as a function of distance from the  center of CenA-MM-Dw3. Also plotted is a range of stream models from \citet{Pearson22}, who explored dynamical models that could reproduce the stream morphology and velocity of the central star cluster associated with CenA-MM-Dw3. While the models leave some freedom in the best fit orbital parameters of the stream, they are representative of the family of models that best match the existing data, and predict the stream velocity gradients plotted.  Note, however, that the sign of the velocity gradient would flip if Dw3 was instead located in front of NGC~5128; the plotted set of models assume that Dw3 is at a slightly larger distance from us than NGC~5128. Also, three of the GCs we associated with CenA-MM-Dw3 and its tidal tail are beyond what \citet{Pearson22} considered the extent of the stream.

This comparison suggests that at most three GCs are associated with CenA-MM-Dw3: the central nuclear star cluster and two GCs in the tidal tails. In this case the GCs would generally follow the predicted velocity gradient, and the other objects would be unassociated with the dwarf. It is likely not tenable for all the GCs to have originated from the dwarf: while \citet{Pearson22} does not make a specific prediction for the velocity dispersion at specific distances along the stream, the dispersion should reflect that of the progenitor and is expected to be low. For example, the intrinsic velocity dispersion along the tidal tails of the Sgr dwarf is only $\sim 15$ km s$^{-1}$ (e.g., \citealt{2020MNRAS.497.4162V}), and the luminosity and other characteristics of CenA-MM-Dw3 are consistent with it having had a Sgr-like progenitor. Hence deviations of at most $\sim 30$--35 km s$^{-1}$ from the mean stream velocity at a given radius would be expected, compared to the observed deviations of $> 100$ km s$^{-1}$ that would be needed to accommodate all of the candidate stream GCs as truly associated with CenA-MM-Dw3. It is also plausible that even fewer than three GCs are linked to the dwarf.  
Given its importance for constraining the properties of the dark matter halo of NGC 5128, searching for additional GCs that could be associated with CenA-MM-Dw3 (i.e. fainter systems) and critically assessing the existing candidates in terms of (for example) chemical abundances should be a high priority moving forward.

\subsection{Second Stream}

The ``second stream" is a stellar feature seen in resolved RGB stars that is
directly south of CenA-MM-Dw3, but much less luminous \citep{Crnojevic16}. It appears to lack a remnant central concentration of stars, suggesting the progenitor dwarf has been fully disrupted. Here we identify two GCs with similar velocities that lie projected on top of the second stream, and have velocities within 60 km s$^{-1}$ of each other. Initial simulations were able to reproduce the general stream morphology using the central GC velocity as an initial condition \citep{Pearson22}, and future refinements may provide tighter constraint on the properties of NGC~5128's dark matter halo in conjunction with the Dw3 stream.  

\subsection{Cloud S}
Another stellar substructure identified by \citet{Crnojevic16} is ``Cloud S", a large, diffuse structure south of the main body of NGC 5128. The boundaries of Cloud S are somewhat difficult to define, but there are at least six GCs that lie in projection on this structure (Figure \ref{fig:marked_map}). These GCs have a mean velocity of $524$ km s$^{-1}$ and a dispersion of only $19$ km s$^{-1}$ (the low dispersion is apparent in Figure \ref{fig:vel_dist_substruc}), suggesting a low mass progenitor. If the single highest velocity is removed as a possible interloper, the velocity dispersion drops to only 13 km s$^{-1}$, consistent with an unresolved intrinsic dispersion given the typical velocity uncertainties for these GCs. In either case, the low dispersion suggests that the association between the GCs and Cloud S could be real.

We note that the boundaries of Cloud S are poorly defined: it is not clear whether there is a single structure from one accretion event or multiple overlapping substructures. An extension of the stellar RGB map to the west of Cloud S could allow better mapping of these substructures and an improved judgement about the association of GCs with Cloud S.

%%%%%%%%%%%%%%%%%%%%%%%%%%%%%%%%%%%%%%%%%%%%%%%%%%%%%%%%%%%%%%%%
\section{Kinematics and Dynamics of NGC 5128}\label{sec:total_mass}

There are a broad range of efforts in the existing literature to provide mass estimators for pressure-supported systems (e.g., \citealt{1981ApJ...244..805B,1986AJ.....92...72R,2008Natur.454.1096S,2010MNRAS.406.1220W}). The goal of these estimators---all of which are essentially variations on the virial theorem---is to allow basic mass measurements that do not require complex dynamical modeling. These estimators all use simplifying assumptions about the spatial distribution and orbital properties of the tracer population, leading to various systematic uncertainties in the resulting masses.

Here we make a hybrid mass estimate of NGC 5128, combining two different estimators that account for both the pressure supported ($M_p$) and rotationally supported ($M_r$) mass components of NGC 5128, summing these to get the total mass. A similar kinematic analysis of NGC~5128 was performed by \citetalias{Woodley2010a} with the GC population available at the time.  We emphasize that this is a preliminary dynamical analysis that will be supplanted by a more sophisticated model in a future paper.

For the pressure supported component, we use the  ``tracer mass estimator'' \citep{Evans2003}, which is well-suited for tracer populations such as GCs that do not necessarily follow the underlying radial distribution of the dark matter.

This mass $M_p$ is given by: 
\begin{equation}
    M_p = \frac{C}{GN} \sum_i (v_{f_i} - v_{sys} )^2 R_i,
    \label{eqn:mass_pressure}
\end{equation}
where $N$ is the number of objects in the sample, $v_{f_i}$ is the measured radial velocity of the tracer object %$(v_p)$ 
with any rotational amplitude removed, $v_{sys}$ is the systemic velocity, and $R_i$ is the projected galactocentric radius of the tracer object.  We determine the systemic velocity, rotation amplitude, and projected rotation axis for the GC system using the kinematic solution described in Section \ref{sub:rotation}.

The constant $C$ is dependent on the shape of the underlying gravitational potential, the radial distribution of the tracers, and the anisotropy of the system. For this estimate, we assume the GC system is spherical and isotropic, so the value of the constant $C$ is given by 
\begin{equation}
    C = \frac{4 (\alpha + \gamma) (4 - \alpha - \gamma) (1 - (\frac{r_{in}}{r_{out}})^{(3-\gamma)})} {\pi (3-\gamma)(1 - (\frac{r_{in}}{r_{out}})^{(4 - \alpha - \gamma)} )},
    \label{eqn:const_c}
\end{equation}
where $r_{in}$ and $r_{out}$ are the three-dimensional radii corresponding to the two-dimensional projected radii $R_{in}$ and $R_{out}$ of the innermost and outermost tracers in the sample.  Because our GC tracers are derived from a wide-angle survey in which the population is traced out to large radii, we assume that $r_{out} \approx R_{out}$ and $r_{in} \approx R_{in}$. We assume an isothermal halo over the radial range of our data, which corresponds to $\alpha = 0$.
The parameter $\gamma$ is calculated from the surface density of the tracer population, and we adopt the values derived in Section \ref{sec:updated_color}.

The mass supported by rotation is determined from the rotational component of the Jeans equation, 
\begin{equation}
    M_r = \frac{R_{out}v_{max}^2}{G},
\end{equation}
where $R_{out}$ is again the projected radius of the outermost tracer in the sample and $v_{max}$ is the rotation amplitude.  We assume here that the rotation of the halo GCs occurs only on simple circular orbits.

Values for rotation parameters and mass estimates are shown in Table \ref{table:mass} and discussed in the following sections.

\begin{figure}[t]
\includegraphics[width=1.0\linewidth]{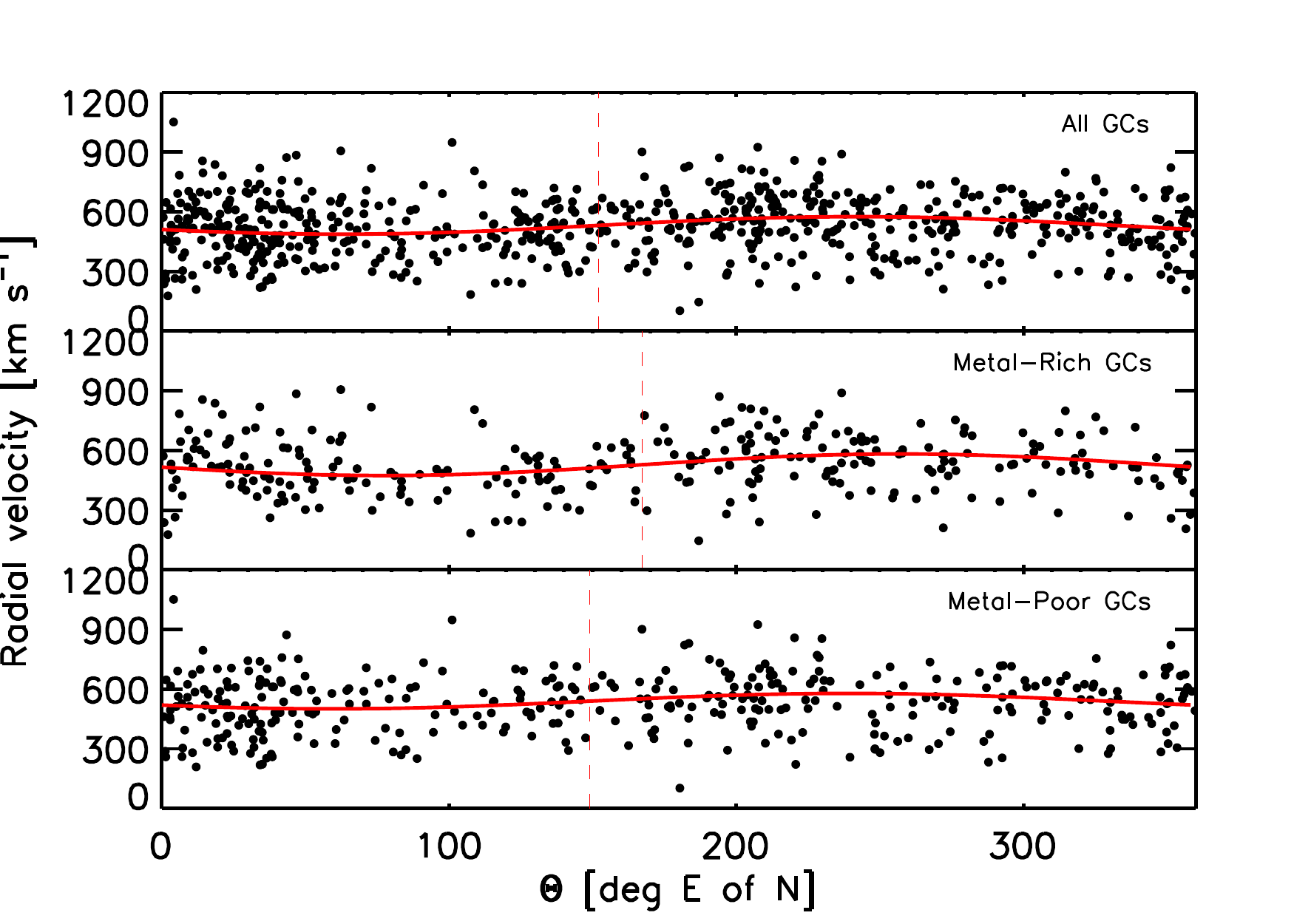}
\caption{Position angle and measured radial velocity are shown for the entire GC sample (top), the metal-rich sample (middle), and the metal-poor sample (bottom).  Each GC sample has been fit with Equation \ref{eqn:rotation} which is overplotted as the red solid curve.  The dashed vertical line denotes the fitted rotation axis of each sample.  The fitted parameters are listed in Table \ref{table:mass}. 
}
\label{fig:rotation}
\end{figure}

\subsection{Rotation Parameters}\label{sub:rotation}

To determine the rotation parameters for the updated GC population, we use the equation
\begin{equation}
    v_p = v_{sys} + \Omega R \sin (\Theta - \Theta_0)
    \label{eqn:rotation}
\end{equation}
described in \citet{Cote2001}, where $v_p$ is the measured radial velocity, where $v_{sys}$ is the systemic velocity, $\Omega R$ is the rotation amplitude (amount of rotation), $\Theta$ is the measured angular position on the projected sky in degrees East of North, and $\Theta_0$ is the rotation axis in the plane of the sky, also in degrees East of North.
The assumptions made include (i) that the GC system of NGC~5128 can be approximated as spherical, (ii) its angular velocity field is constant on spheres, and (iii) its rotation axis lies exactly on the plane of the sky.
The individual GCs are weighted according to their individual radial velocity uncertainties, where the radial velocities and associated uncertainties used are the weighted averages of all previous measurements, as listed in Tables~\ref{table:known_gcs} and \ref{table:new_gcs}. 

Our fitting results for the entire GC sample as well as the metal-rich and metal-poor subpopulations are shown in Figure  \ref{fig:rotation}.  The best fit sine curves clearly show the large dispersion of the GC system with a small rotational component.
The rotation parameters are listed for subsets of the GC population in Table \ref{table:mass},  where errors are calculated using a Monte Carlo bootstrapping method (resampling with replacement).

By comparing the rotation parameters for the sample of all GCs to the sample of GCs unassociated with substructures in the halo, we checked that the substructure-associated GCs do not greatly affect the rotation parameters of the system as a whole.  

Stronger differences can be seen when comparing the rotation parameters of the metal-poor and metal-rich GCs. 
The metal-rich GC subpopulation has a larger rotation amplitude than the metal-poor GC subpopulation. 
Both metallicity subpopulations appear to rotate around the isophotal major axis of the galaxy within $\sim 15$\arcmin. Similar kinematic trends were seen in the  kinematic analysis by \citet{Woodley2010a}, based on a smaller sample of NGC 5128 GCs that extended out to 45\arcmin\ in galactocentric radius. 
Comparing to other galaxies, some evidence for rotation around the photometric major axis has previously been observed for GCs over some radial ranges in the giant ellipticals NGC 4472 \citep{2000AJ....120.2928Z,2003ApJ...591..850C} and M87
\citep{Cote2001,Strader2011}.

\subsection{Mass results}

We determine the rotation and pressure supported masses in cumulative bins of increasing radii.  The results are tabulated in Table \ref{table:mass} with columns displaying the GC subpopulation, the outer radial range of the included GCs in arcminutes, the number of GCs per radial bin, the systemic velocity, the rotation amplitude, the rotation axis, the pressure supported mass, the rotation supported mass, the total mass enclosed, and finally the outer radial range of the included GCs in kpc. Mass errors are calculated using a Monte Carlo bootstrapping method. 
We use an inner radial limit of 5\arcmin\ for all tabulated results. 
As a small number of GCs do not have photometry available, they cannot be classified by subpopulation, causing the number of metal-poor and metal-rich GCs in a given radius range to be less than the number listed in the corresponding row of the ``All" range.

An important issue is that the tracer mass estimators assumes a well-behaved, virialized population, but some of the outer halo GCs appear to be spatially associated with cold stellar debris, leading to both biased and correlated velocities. Therefore, we list values for the total mass with (``All") and without (``Field") the GCs that are potentially associated with substructure in the halo of NGC~5128.

We made two additional assumptions that lead to systematic uncertainties in our masses: that the GC orbits are isotropic, and that the whole dark matter halo is isothermal. These systematic uncertainties are not incorporated into the formal
mass uncertainties. \citet{Evans2003} report a typical systematic uncertainty due to anisotropy of about 30\%. For substantial deviations from an isothermal halo profile, the masses could differ by 40--50\%, though such large variations around the halo scale radius might be surprising.

Table \ref{table:mass} shows that at all radii and for all samples, rotational support is not important compared to pressure support, contributing at most a few percent to our GC-based mass estimates. It also shows that the metal-rich and metal-poor subpopulations generally give consistent mass estimates within their respective uncertainties.

We briefly highlight the masses 
inferred from our analysis at a few specific radii. First, we find a $R<20\arcmin$ (21.6 kpc) mass of $5.1\pm0.6 \times 10^{11} M_{\odot}$. This radius is likely not too far from the dark matter halo scale radius (which is not directly constrained in this modeling). Next, within $R<60\arcmin$ ($\sim 65$ kpc) the enclosed mass is $1.2 \times 10^{12} M_{\odot}$ for both the ``all" sample and ``field" sample of GCs.
This is the largest radius at which the GC population appears dominated by a relaxed, mostly virialized population, though some GCs associated with substructure are present even at these radii. 

Finally, using the ``field"  GCs unassociated with kinematically cold substructure,  we find the total mass enclosed within 110\arcmin~ ($\approx$ 120 kpc) from the center of NGC~5128 to be $(2.5\pm 0.3) \times 10^{12}~ M_\odot$. This is around where the sampling peters out: if one calculates the mass extending to 
the most remote known NGC~5128 cluster at 175\arcmin~ ($\approx 190$ kpc), the value is $(3.2\pm 0.6) \times 10^{12}~ M_\odot$, likely nearing the virial radius of the galaxy. But there are only two objects in the sample between 110\arcmin~ and 175\arcmin, and tracers at the largest radii are likely not in equilibrium, so this ``virial" measurement is primarily an extrapolation of the $\sim$ 120 kpc value.

\subsection{Comparison to previous measurements}\label{sec:compare_mass}

\begin{figure}[t]
\includegraphics[width=1.0\linewidth]{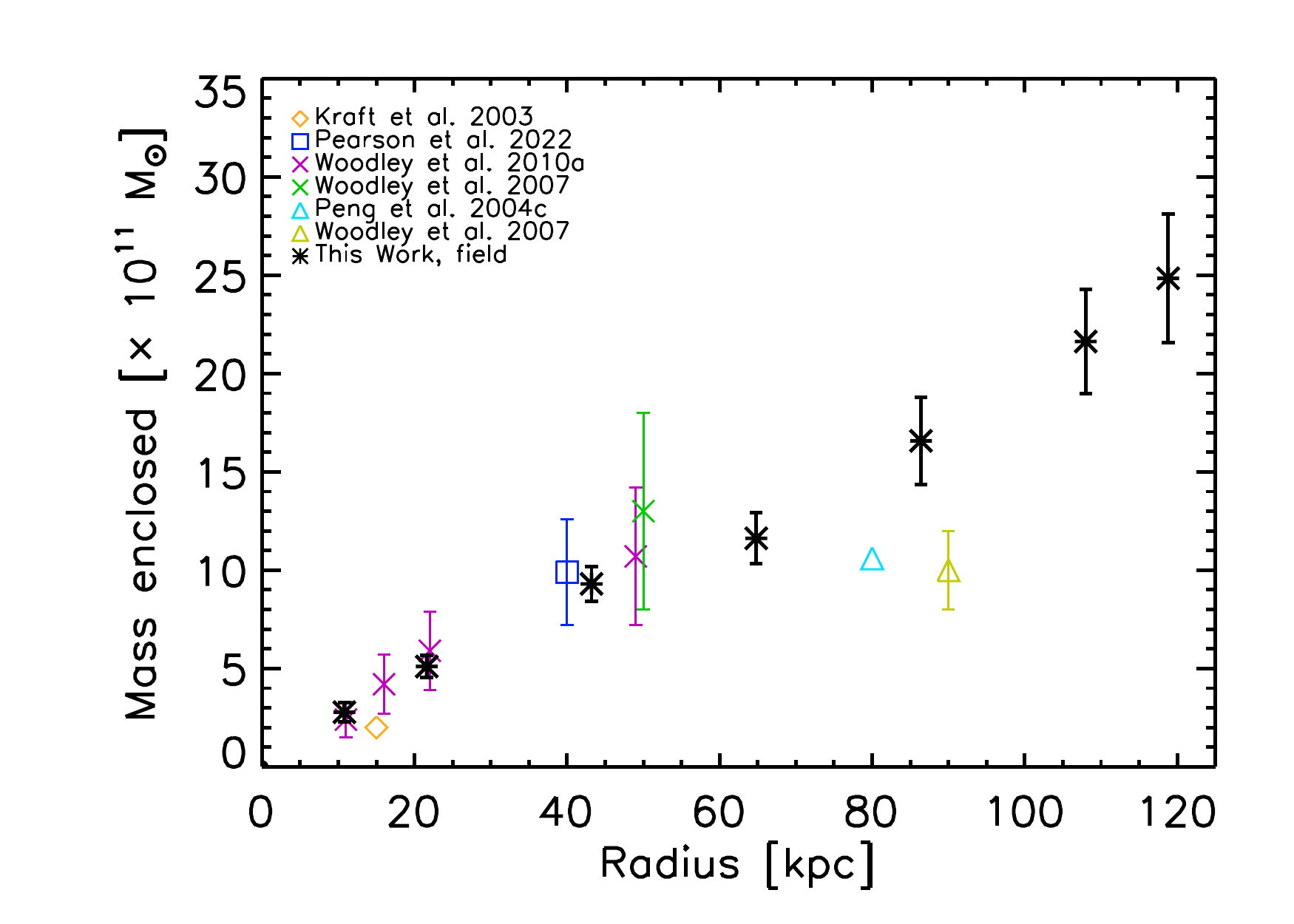}
\caption{Enclosed mass estimates for NGC~5128 using different techniques, expressed in $10^{11} ~M_\odot$.  The method for each technique is listed in Table~\ref{table:mass_comparison}. 
}
\label{fig:mass_comparison}
\end{figure}

There are a number of mass estimates found in the literature that sample a similar spatial range as to our work here, listed in Table~\ref{table:mass_comparison} and plotted in Figure~\ref{fig:mass_comparison}. 

\cite{Peng2004gc2} used 215 GCs extending out to 40 kpc ($\sim 36$\arcmin) to estimate a pressure supported mass of $7.5 \times 10^{11} ~M_\odot$ (see \citealt{Woodley2007} for this mass, corrected from \citealt{Peng2004gc2}).  
We re-evaluate the pressure supported mass using the newly updated sample of GCs out to 36\arcmin~ and obtain $M_p = 8.4 \pm 0.8 \times 10^{11} ~M_\odot$.
\cite{Woodley2007} and \cite{Woodley2010a} measured $M_t = ( 13 \pm 5) \times 10^{11} ~M_\odot$ within 50 kpc and $M_t = ( 10.7 \pm 3.5) \times 10^{11} ~M_\odot$ within 45\arcmin~(49 kpc). 
While not directly tabulated in Table \ref{table:mass}, we evaluate our total mass estimate within 50 kpc and find $M_t = ( 10.5 \pm 1.0) \times 10^{11} ~M_\odot$. 
These mass estimates match well within measurement uncertainties.

Based on X-ray emission from the interstellar medium and the inner radio lobes, \cite{Kraft2003} find that within 15 kpc (13.8\arcmin) of the nucleus, the total mass of NGC~5128 is $\sim 2 \times 10^{11} ~M_\odot$. Within this same projected radius, we find $M_t = ( 3.6 \pm 0.4) \times 10^{11} ~M_\odot$.

\cite{Peng2004PNe} measured the total mass of NGC~5128 using planetary nebulae to be $10.6 \times 10^{11} ~M_\odot$ within 80 kpc (see \citealt{Woodley2007} for this mass, corrected from \citealt{Peng2004gc2}). \cite{Woodley2007} similarly found $M_t = (10\pm2) \times 10^{11} ~M_\odot$ within 90 kpc.
All the GCs in our sample at radii between 80--90 kpc (74--83\arcmin) are located within kinematically cold substructure.  So our closest comparable measurement is $M_t = ( 16.6 \pm 2.2) \times 10^{11} ~M_\odot$ within 80\arcmin, based on the ``farthest" field GC at 71\arcmin~ (76 kpc).

Most recently, \citet{Pearson22} used the morphology of Dw3 and its stellar stream in conjunction with the velocity of the central Dw3 star cluster to constrain the mass of NGC~5128 assuming an NFW dark matter halo profile.  Given this, we list the range of masses allowed at a fiducial radius of 40 kpc in these models in Table~\ref{table:mass_comparison}, $M_t = ( 9.9 \pm 2.7) \times 10^{11} ~M_\odot$, which is in excellent agreement with our GC measurements.

Overall, we find reasonable to excellent agreement with previous masses based on GCs, but tend to find higher masses than previous estimates based on planetary nebulae, likely due to differing anisotropies between these populations not captured in the assumptions of this initial simple modeling.

%%%%%%%%%%%%%%%%%%%%%%%%%%%%%%%%%%%%%%%%%%%%%%%%%%%%%%%%%%%%%%%%%%%%%%%%%%%%%%%
% Summary & Conclusions
%%%%%%%%%%%%%%%%%%%%%%%%%%%%%%%%%%%%%%%%%%%%%%%%%%%%%%%%%%%%%%%%%%%%%%%%%%%%%%%
\section{Summary}\label{sec:conclude3}

We present new radial velocity measurements for 174 previously known and 122 newly confirmed GCs in NGC~5128.  Our spectroscopy was taken between 2017--2022 with Magellan/M2FS and AAT/AAOmega.  We discovered 69 new GCs within 30\arcmin~ and 28 new GCs beyond 50\arcmin, showcasing the continued need for spectroscopy of GC candidates at all radii.  We found that 27\% of priority sample GC candidates were confirmed to be true GCs, compared to only 7\% of those that were not in the priority sample; the fraction of true GCs in the priority sample soars to 68\% within 30' of NGC~5128. 
The population of confirmed GCs now extends out to a projected radius of nearly 190 kpc. This enables us to explore the kinematic properties of the GCs located in the outermost halo of NGC~5128 for the first time.

We test the fidelity of our velocity measurements by comparing to the catalogs of \cite{Woodley2010a} and find excellent agreement, with measurements for nearly all individual GCs within measurement uncertainties.  We remeasured velocities for four GCs with conflicting measurements that were listed in \citetalias{Hughes2021}, and find that with new radial velocity measurements, they can be more confidently classified as GCs.  

Using the measurements presented here, we demonstrate that various discrete groups of GCs projecting onto the most luminous halo streams and overdensities exhibit clear kinematic patterns.  This indicates that a substantial fraction of the outer halo GC population consists of objects accreted along with their now-defunct host galaxies, as has been shown for M31 \citep{Mackey2010, Veljanoski2014}.
A striking feature of many of the ensembles we considered is the coldness of their kinematics, most clearly shown by the low velocity dispersions of GCs associated with CenA-Dw1 and Cloud S, but potentially also with the CenA-MM-Dw3 stream. Definitive proof of these associations will require additional follow-up observations and more modeling.

Based on color, we divide the confirmed GCs into metal-poor ($(g-r)_0 < 0.65$) and metal-rich subpopulations.
We find that the radial distribution of the metal-rich subpopulation has a steeper slope and is more centrally concentrated, which follows trends seen in other extragalactic studies (e.g., \citealt{Brodie06, Faifer2011, Forbes2012}). 
The metal-poor population is more radially extended, and there are more metal-poor GCs than metal-rich GCs beyond $\sim 9\arcmin$.  Additionally, $\approx 80$\% of the GCs associated with halo substructure are metal-poor, providing evidence for the ongoing buildup of its halo via the accretion of lower-mass galaxies.

We find an overall rotation signature of the GC population at all radii. This rotation is not driven by GCs clearly associated with underlying substructure. Comparing the rotation of the metal-poor and metal-rich GC subpopulations, we find while they have similar projected rotation axes, the metal-rich GCs have a larger rotation amplitude than the metal-poor GCs at all radii.
This is in contrast with observations in the Milky Way where the halo GC population appears to exhibit at most only a mild net rotation (e.g. \citealt{Harris2001, Brodie06}), though in M31 the metal-poor GCs have significant rotation, albeit at half the rate of the most metal-rich clusters \citep{2016ApJ...824...42C}.

We use the GC velocities to estimate the enclosed mass of NGC~5128 at a range of radii using the tracer mass estimator. We find a dynamical mass of $1.2\pm0.1 \times 10^{12} M_{\odot}$ ($2.5\pm0.3 \times 10^{12} M_{\odot}$) within a radius of $\sim 65$ kpc ($\sim 120$ kpc), but this result is preliminary, and we will present detailed dynamical modeling in a follow-up work (Dumont et al., in prep).

%-------------------------------------------------------------------
% Acknowledgements
%-------------------------------------------------------------------

\acknowledgments

Based in part on data acquired at the Anglo-Australian Telescope, under programs A/2017A/01 and A/2022A/01 (PI: D. Forbes). We acknowledge the traditional custodians of the land on which the AAT stands, the Gamilaraay people, and pay our respects to elders past and present.  Based on observations at the Anglo-Australian Telescope under community access time granted by NOIRLab (NOIRLab Prop. ID 2019A-0157, 2021A-0252; PI: A. Hughes and Prop. ID 2017A-0305; PI. D. Sand).  NOIRLab is managed by the Association of Universities for Research in Astronomy (AURA) under a cooperative agreement with the National Science Foundation.

AKH and DJS acknowledge support from NSF grants AST-1821967 and AST-1813708. ACS acknowledges support from NSF grant AST-1813609. JS acknowledges support from NSF grants AST-1514763 and AST-1812856 and from the Packard Foundation. NC is partially supported by NSF grant AST-1812461. Research by DC is supported by NSF grant AST-1814208.  MM acknowledges support from NSF grants AST-0923160 and AST-1815403.

\vspace{5mm}
\facilities{Magellan:Clay (Megacam, M2FS), AAT (2dF/AAOmega) }

\software{Source Extractor \citep{Bertin96},  SWarp \citep{Bertin2010}, The IDL Astronomy User's Library \citep{IDLforever}, {\sc 2dfdr}    }

%BIBLIOGRAPHY / REFERENCES
\bibliographystyle{apj}
\bibliography{thepaper}

    \appendix \label{appendix}
 \section{Discussion of Conflicting Velocity Measurements}

In this Appendix, we discuss several types of ``conflicting" NGC~5128 GC velocity measurements.  The first correspond to new NGC~5128 GC measurements which disagree with the weighted radial velocity measurements reported in \citetalias{Woodley2010a} by $> 2\sigma$, and can be seen as outliers in Figure~\ref{fig:vel_comp}. The second corresponds to velocity measurements in the literature of the same GC which are in conflict with each other, and which we were able to resolve by obtaining a new velocity measurement. In the final case, the reported  velocities (either in the literature or here) could be consistent with a GC, but Gaia DR3 provides evidence that the source is a foreground star.

In all cases of GCs with discrepant measurements, we visually checked the PISCeS data for close companions to rule out target misidentification, but found no viable interlopers.

\subsection{Outlier New GC and W10 Measurements} 

\cite{Woodley2010a} find H21-217604 ($\alpha= 201.15733^{\circ}$, $\delta=-43.27402^{\circ}$) to have a velocity
of $212\pm10$ km s$^{-1}$. We have two independent new velocity measurements of this object: $649\pm18$ km s$^{-1}$ (from M2FS) and $679\pm18$ km s$^{-1}$ (from AAOmega). The agreement between these two new measurements provides strong evidence that they are correct, and adopted the weighted average of these, excluding the \cite{Woodley2010a} value.

For four additional objects, we have a new M2FS spectrum that gives a velocity inconsistent with the published one, discussed in detail below.

\cite{Woodley2010a} find a velocity of $189\pm34$ km s$^{-1}$ for H21-210609 ($\alpha=201.08215^{\circ}$, $\delta=-42.84548^{\circ}$). Our M2FS velocity is $691\pm42$  km s$^{-1}$, with the H$\beta$ and Mg$b$ lines both clearly seen in the spectrum and both giving this same velocity within the uncertainties. With no specific reason to doubt our new, NGC 5128--consistent velocity, we adopt it and exclude the \cite{Woodley2010a} value.

\cite{Woodley2010a} find a velocity of $926\pm44$ km s$^{-1}$ for H21-217335 ($\alpha=201.15402^{\circ}$, $\delta= -43.30883^{\circ}$). Our M2FS velocity is $566\pm43$  km s$^{-1}$. Both H$\beta$ and Mg$b$ lines agree with this same M2FS velocity and both are seen clearly in the good S/N spectrum. Given this agreement, we adopt the new M2FS velocity with confidence.

\cite{Woodley2010a} find a velocity of $183\pm49$ km s$^{-1}$ for H21-217458 ($\alpha=201.15564^{\circ}$, $\delta=-43.10869^{\circ}$). Our M2FS velocity is $670\pm55$  km s$^{-1}$. Both H$\beta$ and Mg$b$ lines agree with this same velocity, though the Mg$b$ lines are seen more clearly than H$\beta$, which is weaker. Given the clear velocity from the Mg$b$ lines, we adopt our new M2FS measurement, though with slightly lower confidence.

\cite{Woodley2010a} find a velocity of $297\pm21$ km s$^{-1}$ for H21-216619 ($\alpha=201.14640^{\circ}$, $\delta=-42.96653^{\circ}$). Our M2FS velocity is $719\pm73$ km s$^{-1}$. This is a metal-poor GC and we do not get a clean velocity measurement from the Mg$b$ lines---there is only a clear H$\beta$ velocity. All ancillary information suggests this source is indeed a confirmed GC, and we adopt our M2FS measurement as its final velocity. However, we have the lowest confidence in this object among all of the conflicting objects, and it should be a  priority for re-measurement in the future.

\subsection{Resolved Conflicting Literature Measurements}

We measured velocities for four GCs listed in the table of previously classified objects with conflicting measurements from \citetalias{Hughes2021} (Table~6 in the Appendix of that work). Our new measurements agree with at least one of their previous measurements in the literature, and they are within the velocity range expected for GCs in NGC~5128, so we include these four GCs in Table~\ref{table:known_gcs} with a weighted velocity measurement that excludes the inconsistent measurement from the weighted average.

The first, H21-221818 at position    
$\alpha$=$201.20335^{\circ}$, $\delta$=$-43.27145^{\circ}$, was noted as a GC by \cite{Woodley2010a} with a velocity of $426\pm41$ km~s$^{-1}$ and as a background galaxy by \cite{Beasley2008} with a velocity of 42000 km~s$^{-1}$.  We measure it to have a velocity of $440\pm24$ km~s$^{-1}$ using M2FS, confirming it to be a GC.
The second, H21-257878 at position  
$\alpha$=$201.59204^{\circ}$, $\delta$=$-43.15295^{\circ}$, was noted as a GC by \cite{Woodley2005} with a velocity of $505\pm78$ km~s$^{-1}$ and as a background galaxy by \cite{Beasley2008} with a velocity of 21000 km~s$^{-1}$.  We measure it to have a velocity of $447\pm56$ km~s$^{-1}$ using AAOmega, confirming it to be a GC.
The third, at position
$\alpha$=$201.54395^{\circ}$, $\delta$=$-43.01830^{\circ}$, was noted as a foreground star by \cite{Beasley2008} with a velocity of $231\pm143$ km~s$^{-1}$ and a GC by \cite{Woodley2010a} with a velocity of $311\pm81$.  We measure it to have a velocity of $449\pm36$ km~s$^{-1}$ using AAOmega, confirming it to be a GC (owing to the large uncertainties on the literature measurements, the conflict among the velocities here was only marginal).
It did not receive an ID in \citetalias{Hughes2021}, and we label it H22-459 in Table~\ref{table:known_gcs}.  The fourth, H21-265571 at position
$\alpha$=$201.66045^{\circ}$, $\delta$=$-42.76268^{\circ}$, was noted as a GC with radial velocity measurements of $474\pm65$ km s$^{-1}$ in \citet{Harris2002},
$492\pm37$ km~s$^{-1}$ in \citet{Woodley2007}, and $627\pm21$ km~s$^{-1}$ in \citet{Woodley2010a}. We measure it to have a velocity of $602\pm15$ km~s$^{-1}$ using AAOmega, confirming it to be a GC.

\subsection{Resolving Conflicting Gaia and Velocity Evidence}

As noted in Section~\ref{subsec:confirmed_clusters}, we used a $v_r > 250$~km~s$^{-1}$ threshold for deciding on clusters being confirmed GCs based on their velocity alone.  This was based at least in part on the large number of GC candidates below this velocity that turned out to be stars with highly significant ($>$5$\sigma$) proper motions in Gaia DR3.  However, even above this velocity, there were nine new objects and four literature objects with velocities in the range 250--350~km~s$^{-1}$ and $>$5$\sigma$ Gaia DR3 proper motions.  The nine new objects can be found listed amongst the stars in Table~\ref{table:bad}.  The four previous literature objects at $v_r > 250$~km~s$^{-1}$ with $>$5$\sigma$ proper motion detections are H21-257155 ($\alpha$=$201.58596^{\circ}$, $\delta$=$-42.89608^{\circ}$), GC0333 ($\alpha$=$201.48435^{\circ}$, $\delta$=$-43.02578^{\circ}$), H21-247489 ($\alpha$=$201.48857^{\circ}$, $\delta$=$-43.68582^{\circ}$), and H21-202595 ($\alpha$=$200.99837^{\circ}$, $\delta$=$-42.92202^{\circ}$). These can all confidently be classified as stars.

Only one object with a Gaia proper motion had a velocity $> 350$ km s$^{-1}$ that would be very unusual for a foreground star: \citet{Beasley2008} report the object H21-247489 ($\alpha=201.48857^{\circ}$, $\delta=-43.68581^{\circ}$) 
to have a radial velocity of $601\pm66$ km s$^{-1}$, but this source also has a high-significance ($>5\sigma$) Gaia DR3 proper motion and does not appear extended in ground-based imaging or in Gaia. Pending an additional radial velocity measurement, we exclude this from the catalog of velocity-confirmed GCs.  

On the other hand, we keep one object that has a high significance Gaia DR3 proper motion: H21-335694  ($\alpha$=$202.33212^{\circ}$, $\delta$=$-41.67356^{\circ}$).  This object has a high S/N spectrum with all absorption lines consistent with its measured velocity at $v_r$=452~km~s$^{-1}$ and appears extended in our ground-based imaging.  In Gaia DR3, this object shows a very high proper motion $\mu_\alpha=-12.586\pm0.113$ and $\mu_\delta=-0.046\pm0.097$, but only in the $\alpha$ direction, and also has no measurable parallax, which is unusual for a high proper motion object with its $G$ mag. We suspect the Gaia measurement may be erroneous and keep this object, pending an update of the Gaia measurement in a future data release.

\newpage

\begin{table*}[t]
\centering
\caption{M2FS observation summary}
\begin{tabular}{c l c c c r}
\hline \hline
Field & UT Date & \multicolumn{2}{c}{Field Center} & Time  \\
 &  & R.A. (J2000) & Decl. (J2000) &   (hr) \\
\hline
1 & 2017 Feb 25  & 13:25:41.1  & $-$43:22:13 & 3.0 \\ 
2 & 2017 May 21  & 13:23:53.0  & $-$42:59:41 & 3.0 \\ 
3 & 2017 May 21 & 13:28:11.4  & $-$43:21:34 & 3.0 \\ 
4 & 2017 June 4 & 13:24:07.6  & $-$42:35:50 & 3.2 \\ 
5 & 2018 May 12  & 13:30:31.1  & $-$42:03:14 & 2.5 \\ 
6 & 2019 Feb 27  & 13:25:10.3  & $-$42:56:29 & 3.3 \\ 
7 & 2019 March 7 & 13:26:51.9  & $-$42:39:16 & 1.5 \\ 
\multirow{2}{*}{8} & 2019 March 5 & \multirow{2}{*}{13:24:53.8} & \multirow{2}{*}{$-$43:27:51} & 1.75  \\ 
& 2019 June 2 &  &  & 1.5   \\ 
\hline
\end{tabular}
\label{table:m2fs_observing}
\end{table*}

\begin{table*}[t]
\centering
\caption{AAOmega observation summary}
\medskip
\begin{tabular}{c l c c c l}
\hline \hline
Field & UT Date & \multicolumn{2}{c}{Field Center} & Time  \\
 &  & R.A. (J2000) & Decl. (J2000) &   (hr) \\
\hline
9 & 2017 May 4  & 13:26:09 & $-$42:53:02 & 4.5  \\ 
10 & 2017 May 5 & 13:27:14 & $-$42:26:54  & 6.5  \\ 

11 & 2019 March 30 & 13:29:28 & $-$43:47:49 & 2.0  \\

12 & 2019 March 30 & 13:21:40 & $-$43:46:58 & 2.5 \\ 

13 & 2022 April 23 & 13:31:30 &  $-$42:43:09 & 4.0 \\ 

14 & 2022 May 5 & 13:23:19  & $-$41:44:00 & 6.5 \\ 
\hline
\end{tabular}
\label{table:2df_observing}
\end{table*}

\begin{table*}[t]
\small
\centering
\caption{Radial velocity measurements of previously known GCs in NGC 5128}
\setlength{\tabcolsep}{3pt}

\begin{tabular}{c c c c c c c c c c c}
\hline \hline
H21/22-ID & R.A. & Decl. & g & r &  (g-r)$_0$ & (u-z)$_0$ & M2FS v$_r$  & AAOmega v$_r$ & weighted v$_r$ & Notes \\
  & (deg J2000) & (deg J2000) &   (mag) & (mag) & (mag) & (mag) & (km/s) & (km/s) & (km/s) &  \\
\hline
H21-155942 & 200.50859 & -42.53533 & 18.95 & 18.25 & 0.54 & 2.14 &  ...  & 643$\pm$10 & 639$\pm$2 & 12 \\
H21-194226 & 200.90970 & -42.77302 & 18.86 & 18.22 & 0.49 & 1.99 &  ...  & 479$\pm$26 & 486$\pm$20 & 7 \\
H21-195812 & 200.92642 & -43.16045 & 19.37 & 18.64 & 0.59 & 2.36 &  ...  & 374$\pm$17 & 379$\pm$15 & 4,7 \\
H21-196535 & 200.93407 & -43.18659 & 18.12 & 17.38 & 0.74 & ...  &  ...  & 647$\pm$17 & 648$\pm$6 & 1,2,4,7,9,10 \\
H21-196891 & 200.93760 & -43.01983 & 19.67 & 18.85 & 0.69 & 2.77 & 589$\pm$14 &  ...  & 581$\pm$13 & 9 \\
H21-198648 & 200.95671 & -43.24220 & 19.38 & 18.55 & 0.68 & 2.71 &  ...  & 675$\pm$17 & 682$\pm$15 & 4,7 \\
H21-200443 & 200.97577 & -43.36572 & 22.67 & 21.65 & 1.01 & ...  & 459$\pm$44 &  ...  & 470$\pm$39 & 9 \\
\hline
\end{tabular}
\label{table:known_gcs}
\begin{tablenotes}
      \footnotesize
      \item The complete table will be available online.
      \item Notes: Radial velocity measured by 
    1 = \citet{VanDenBergh1981}. 
    2 = \citet{Hesser1986}. 
    3 = \citet{Harris1992}. 
    4 = \citet{Peng2004gc1}. 
    5 = \citet{Woodley2005}. 
    6 = \citet{Rejkuba2007}. 
    7 = \citet{Beasley2008}. 
    8 =  \citet{Woodley2010b}.
    9 =  \citet{Woodley2010a}.
    10 =  \citet{Taylor2010}.
    11 =  \citet{Voggel2020}.
    12 =  \citet{Dumont2021}.
\end{tablenotes}
\end{table*}

\begin{table*}[ht]
\centering
\caption{Radial velocity measurements of newly confirmed GCs in NGC~5128}
\begin{tabular}{c c c c c c c c c c}
\hline \hline
H21/H22-ID & R.A. & Decl. & g & r & (g-r)$_0$ & (u-z)$_0$ & M2FS v$_r$  & AAOmega v$_r$  \\
  & (deg J2000) & (deg J2000) &   (mag) & (mag) & (mag) & (mag) & (km/s) & (km/s)  \\
\hline
H21-082748 & 199.70710 & -43.49976 & 19.94 & 19.41 & 0.37 & 1.56 &  ...  & 300$\pm$18 \\
H21-115063 & 200.10130 & -43.56268 & 19.61 & 18.81 & 0.65 & 2.40 &  ...  & 257$\pm$13 \\
H21-144922 & 200.39422 & -43.51873 & 19.26 & 18.60 & 0.51 & 2.01 &  ...  & 513$\pm$17 \\
H21-155829 & 200.50759 & -43.43897 & 20.42 & 19.45 & 0.81 & 2.82 &  ...  & 536$\pm$22 \\
H21-166444 & 200.61574 & -41.74595 & 21.49 & 20.20 & 1.17 & 3.33 &  ...  & 269$\pm$14 \\
H21-166516 & 200.61649 & -42.08955 & 21.11 & 20.28 & 0.69 & 2.22 &  ...  & 275$\pm$14 \\
H21-184891 & 200.80609 & -43.83047 & 19.34 & 18.63 & 0.56 & 2.21 &  ...  & 502$\pm$15 \\
H21-186688 & 200.82601 & -43.03390 & 20.61 & 19.89 & 0.57 & 2.20 & 651$\pm$32 &  ...  \\
H21-186962 & 200.82928 & -42.83677 & 20.00 & 19.14 & 0.70 & 2.84 & 530$\pm$41 &  ...  \\
\hline
\end{tabular}
\label{table:new_gcs}
\begin{tablenotes}
      \footnotesize
      \item The complete table will be available online.
\end{tablenotes}
\end{table*}

\begin{table*}[t]
\centering
\caption{Foreground stars and background galaxies in the vicinity of NGC 5128}
\begin{tabular}{c c c c c c c}
\hline \hline
H21-ID & R.A. & Decl. & g & r &  weighted v$_r$  & type \\
  & (deg J2000) & (deg J2000) &   (mag) & (mag) & (km/s) &  \\
\hline
H21-045317 & 199.13075 & -43.57504 & 19.97 & 19.04 & 4$\pm$14 & star  \\
H21-048213 & 199.19358 & -43.48774 & 21.45 & 19.76 & -418$\pm$12 & star  \\
H21-048293 & 199.19549 & -43.57120 & 19.32 & 18.46 & 91$\pm$17 & star  \\
H21-049729 & 199.22672 & -43.35489 & 21.18 & 20.52 & -377$\pm$20 & star  \\
H21-054879 & 199.33153 & -43.54923 & 19.90 & 19.29 & 280$\pm$15 & star  \\
H21-055213 & 199.33754 & -43.33671 & 19.58 & 18.24 & -4$\pm$22 & star  \\
H21-055349 & 199.33936 & -43.97305 & 21.18 & 20.12 & -18$\pm$16 & star  \\
H21-059335 & 199.39602 & -43.17051 & 20.21 & 19.29 & -37$\pm$20 & star  \\

\hline
\end{tabular}
\label{table:bad}
\begin{tablenotes}
      \footnotesize
      \item The complete table will be available online.
\end{tablenotes}
\end{table*}

\begin{table*}[t]
\centering
\small
\caption{The mass of NGC~5128}
\begin{tabular}{c c c c c c c c c c}
\hline \hline
GCs & outer $R$      & $N$ & $v_{sys} $   & $\Omega R$    & $\Theta_0$   & $M_p$ & $M_r$ & $M_t$ & outer $R$ \\
    & (arcmin) &    & (km s$^{-1}$) & (km s$^{-1}$) & (deg. E of N) & ($\times 10^{11} M_\odot$)&  ($\times 10^{11} M_\odot$) & ($\times 10^{11} M_\odot$) & (kpc) \\
\hline
All GCs & 10  & 221 & 535$\pm$13 & 48$\pm$37 & 192$\pm$172 & 2.71$\pm$0.45 & 0.063$\pm$0.049 & 2.78$\pm$0.49 & 10.8 \\ 
All GCs & 20  & 387 & 532$\pm$9 & 54$\pm$16 & 175$\pm$32 & 4.97$\pm$0.52 & 0.146$\pm$0.070 & 5.10$\pm$0.57 & 21.6 \\ 
All GCs & 30  & 454 & 538$\pm$8 & 43$\pm$17 & 175$\pm$39 & 6.99$\pm$0.62 & 0.137$\pm$0.076 & 7.14$\pm$0.66 & 32.4 \\ 
All GCs & 40  & 485 & 539$\pm$7 & 48$\pm$13 & 170$\pm$23 & 9.08$\pm$0.82 & 0.231$\pm$0.113 & 9.30$\pm$0.89 & 43.2 \\ 
All GCs & 60  & 504 & 540$\pm$7 & 45$\pm$13 & 169$\pm$20 & 11.82$\pm$1.21 & 0.303$\pm$0.155 & 12.13$\pm$1.31 & 64.8 \\ 
All GCs & 80  & 517 & 537$\pm$7 & 46$\pm$15 & 170$\pm$25 & 17.77$\pm$2.36 & 0.422$\pm$0.211 & 18.19$\pm$2.49 & 86.4 \\ 
All GCs & 100 & 531 & 534$\pm$7 & 48$\pm$15 & 165$\pm$23 & 27.59$\pm$3.67 & 0.553$\pm$0.295 & 28.14$\pm$3.84 & 108.0 \\ 
All GCs & 110 & 533 & 535$\pm$8 & 50$\pm$14 & 165$\pm$23 & 30.71$\pm$4.28 & 0.674$\pm$0.301 & 31.43$\pm$4.44 & 118.8 \\ 
All GCs & 175 & 535 & 536$\pm$7 & 50$\pm$14 & 164$\pm$21 & 38.55$\pm$6.97 & 0.908$\pm$0.492 & 39.56$\pm$7.28 & 189.0 \\

\hline

Field GCs & 60 & 499 & 540$\pm$7 & 47$\pm$12 & 168$\pm$17 & 11.35$\pm$1.22 & 0.305$\pm$0.142 & 11.62$\pm$1.30 & 64.8 \\ 
Field GCs & 80 & 505 & 539$\pm$7 & 47$\pm$13 & 167$\pm$23 & 16.15$\pm$2.09 & 0.416$\pm$0.206 & 16.58$\pm$2.21 & 86.4 \\ 
Field GCs & 100 & 513 & 539$\pm$8 & 46$\pm$14 & 165$\pm$23 & 21.14$\pm$2.53 & 0.505$\pm$0.259 & 21.63$\pm$2.66 & 108.0 \\ 
Field GCs & 110 & 515 & 539$\pm$7 & 48$\pm$14 & 166$\pm$31 & 24.25$\pm$3.14 & 0.608$\pm$0.294 & 24.84$\pm$3.28 & 118.8 \\ 
Field GCs & 175 & 517 & 541$\pm$7 & 47$\pm$12 & 164$\pm$16 & 30.56$\pm$5.29 & 0.826$\pm$0.454 & 31.50$\pm$5.55 & 189.0 \\

\hline
MP GCs & 20 & 202 & 539$\pm$13 & 43$\pm$25 & 169$\pm$69 & 4.86$\pm$0.77 & 0.094$\pm$0.095 & 4.96$\pm$0.83 & 21.6 \\ 
MP GCs & 40 & 264 & 549$\pm$11 & 35$\pm$20 & 168$\pm$62 & 8.67$\pm$1.15 & 0.116$\pm$0.127 & 8.78$\pm$1.23 & 43.2 \\ 
MP GCs & 60 & 280 & 549$\pm$10 & 34$\pm$20 & 165$\pm$71 & 11.63$\pm$1.74 & 0.166$\pm$0.180 & 11.79$\pm$1.86 & 64.8 \\ 
MP GCs & 80 & 290 & 545$\pm$10 & 34$\pm$20 & 163$\pm$68 & 19.85$\pm$3.62 & 0.227$\pm$0.249 & 20.18$\pm$3.76 & 86.4 \\

\hline
MR GCs & 20 & 182 & 533$\pm$13 & 63$\pm$23 & 178$\pm$120 & 6.10$\pm$0.97 & 0.197$\pm$0.102 & 6.30$\pm$1.05 & 21.6 \\ 
MR GCs & 40 & 218 & 533$\pm$11 & 62$\pm$19 & 172$\pm$64 & 11.72$\pm$1.78 & 0.379$\pm$0.186 & 12.12$\pm$1.91 & 43.2 \\ 
MR GCs & 60 & 221 & 532$\pm$11 & 61$\pm$21 & 172$\pm$52 & 12.85$\pm$2.11 & 0.410$\pm$0.218 & 13.26$\pm$2.27 & 64.8 \\ 
MR GCs & 80 & 224 & 532$\pm$11 & 62$\pm$18 & 173$\pm$63 & 18.23$\pm$3.16 & 0.677$\pm$0.331 & 18.93$\pm$3.40 & 86.4 \\

\hline
\end{tabular}
\label{table:mass}
\begin{tablenotes}
      \footnotesize
      \item Field GCs are taken from the `All GCs' sample but with those GCs associated with clear stellar substructures removed.
      \item MP GCs are from the metal poor GC sample defined in Section~\ref{sec:updated_color}.
      \item MR GCs are from the metal rich GC sample defined in Section~\ref{sec:updated_color}.
\end{tablenotes}

\end{table*}

\begin{table*}[t]
\centering
%\small
\caption{Mass estimates for NGC~5128 from recent literature}
%\medskip
%\setlength{\tabcolsep}{4pt}
%\begin{adjustwidth}{-1.1cm}{}
\begin{tabular}{l l l l}
\hline \hline

Reference & $M_t$ ($10^{11} M_\odot$) & $R_{max}$ (kpc) & Method \\
\hline

\cite{Kraft2003} & $\sim2$ & 15  & X-ray emission \\

\cite{Pearson22} & 9.9 $\pm$ 2.7 & 40 & Stream modeling \\

\cite{Woodley2010a} & $10.7\pm3.5$ & 49  & Dynamical tracers: GCs \\ 
\cite{Woodley2007} & $13\pm5$ & 50  & Dynamical tracers: GCs \\
\cite{Peng2004PNe} & 10.6 & 80  & Dynamical tracers: PNe$^1$ \\
\cite{Woodley2007} & $10\pm2$ & 90  & Dynamical tracers: PNe \\
This work & $32 \pm 6$ & 190 & Dynamical tracers: GCs \\
\cite{Karachentsev2007} & 64--81 & 400  & Orbital/virial; Dynamical tracers: dwarf galaxies \\

\hline
\end{tabular}
\label{table:mass_comparison}
\begin{tablenotes}
      \footnotesize
      \item 1. See \cite{Woodley2007} for this mass, corrected from \cite{Peng2004PNe}
\end{tablenotes}
\end{table*}

\end{document}